\documentclass[12pt]{article}
\usepackage[a4paper, total={7in, 9.3in}]{geometry}
\usepackage{amssymb}
\usepackage{amsmath}
\usepackage{graphicx}
\usepackage{mathdots}
\usepackage{slashed}
\usepackage{verbatim}
\usepackage{xcolor}
\usepackage[utf8]{inputenc}

\usepackage{hyperref}

\usepackage{IEEEtrantools}

\def\makeatletter{\catcode`\@=11}
\makeatletter
\def\mathbox#1{\hbox{$\m@th#1$}}%
\def\math@ccstyles#1#2#3#4#5#6#7{{\leavevmode
      \setbox0\mathbox{#6#7}%
      \setbox2\mathbox{#4#5}%
      \dimen@ #3%
      \baselineskip\z@\lineskiplimit#1\lineskip\z@
      \vbox{\ialign{##\crcr
             \hfil \kern #2\box2 \hfil\crcr
             \noalign{\kern\dimen@}%
             \hfil\box0\hfil\crcr}}}}
\def\mathaccstyles{\math@ccstyles\maxdimen}
\def\maththroughstyles{\math@ccstyles{-\maxdimen}}
\def\unity%
 {\maththroughstyles{.45\ht0}\z@\displaystyle {\mathchar"006C}\displaystyle 1}
 
\usepackage{tikz}
\usetikzlibrary{positioning,arrows}
\usetikzlibrary{decorations.pathmorphing,calc}
\usetikzlibrary{decorations.markings}
\newif\ifmirrorsemicircle
\usepackage{float}
\usepackage{ifthen}
\usepackage{enumerate}
\usepackage{amsmath}
\usepackage{amssymb}
\usepackage[mathscr]{eucal}
\usepackage[errorshow]{tracefnt}
\usepackage{amsmath}
\usepackage{slashed}
\usepackage{amssymb}
\usepackage{array}
\usepackage{amsthm}
\usepackage{eucal}
\usepackage{epsf}
\usepackage{epsfig}
\usepackage{psfrag}
\usepackage{multirow}
\usepackage{wasysym} 
\usepackage{tikz-cd} 
\usepgfmodule{shapes}
\usetikzlibrary{backgrounds, arrows,calc,shapes,decorations.pathreplacing, decorations.markings, automata,positioning}
\usepackage{verbatim}

\tikzset{
    wave amplitude/.initial=0.2cm,
    wave count/.initial=8,
    mirror semicircle/.is if=mirrorsemicircle,
    mirror semicircle=false,
    wavy semicircle/.style={
        to path={
            let \p1 = (\tikztostart),
            \p2 = (\tikztotarget),
            \n1 = {veclen(\x2-\x1,\y2-\y1)},
            \n2 = {atan2(\x2-\x1,\y2-\y1))} in
            plot [
                smooth,
                samples=(\pgfkeysvalueof{/tikz/wave count}+0.5)*8+1, 
                domain=0:1,
                shift={($(\p1)!0.5!(\p2)$)}
            ] ({ 
                (\x*180-\n2 + 180 + \ifmirrorsemicircle 1 \else -1 \fi * 90%
            }:{ 
                (%
                    \n1/2+\pgfkeysvalueof{/tikz/wave amplitude} * %
                    sin(
                        \x * 360 * (\pgfkeysvalueof{/tikz/wave count} + 0.5%
                    )%
                )%
            })
        } (\tikztotarget)
    }
}

\tikzset{
    partial ellipse/.style args={#1:#2:#3}{
        insert path={+ (#1:#3) arc (#1:#2:#3)}
    }
}

\pgfdeclaredecoration{complete sines}{initial}
{
    \state{initial}[
        width=+0pt,
        next state=sine,
        persistent precomputation={\pgfmathsetmacro\matchinglength{
            \pgfdecoratedinputsegmentlength / int(\pgfdecoratedinputsegmentlength/\pgfdecorationsegmentlength)}
            \setlength{\pgfdecorationsegmentlength}{\matchinglength pt}
        }] {}
    \state{sine}[width=\pgfdecorationsegmentlength]{
        \pgfpathsine{\pgfpoint{0.25\pgfdecorationsegmentlength}{0.5\pgfdecorationsegmentamplitude}}
        \pgfpathcosine{\pgfpoint{0.25\pgfdecorationsegmentlength}{-0.5\pgfdecorationsegmentamplitude}}
        \pgfpathsine{\pgfpoint{0.25\pgfdecorationsegmentlength}{-0.5\pgfdecorationsegmentamplitude}}
        \pgfpathcosine{\pgfpoint{0.25\pgfdecorationsegmentlength}{0.5\pgfdecorationsegmentamplitude}}
}
    \state{final}{}
}

\tikzset{ 
    fermion/.style={thick, draw=mygreen, postaction={decorate}, decoration={markings, mark=at position .5 with {\arrow[mygreen]{triangle 45}}}} ,
    scalar/.style={thick, draw=blue, postaction={decorate}, decoration={markings, mark=at position .43 with {\arrow[blue]{triangle 45}}}} ,
    scalarr/.style={thick, draw=blue, postaction={decorate}, decoration={markings, mark=at position .56 with {\arrowreversed[blue]{triangle 45}}}},
    photon/.style={decorate, draw=black,thick,
        decoration={complete sines,amplitude=5pt, segment length=7pt}},
         photon1/.style={decorate, draw=black,thick,
        decoration={complete sines,amplitude=3pt, segment length=5pt}},
    photon2/.style={decorate, draw=black,thick,
        decoration={complete sines,amplitude=3pt, segment length=4pt}}
    dashed2/.style={decorate,
    draw=black,thick,decoration={dashed,segment length=7pt}} 
      }

 \numberwithin{equation}{section}

\begin{document}

\begin{flushright}\footnotesize

\texttt{ICCUB-20-008}
\vspace{0.6cm}
\end{flushright}

\mbox{}
\vspace{0truecm}
\linespread{1.1}


\centerline{\LARGE \bf On the UV completion of the $O(N)$ model in }
\medskip

\centerline{\LARGE \bf  $6-\epsilon$ dimensions: a stable large-charge sector}

\medskip


\vspace{.4cm}

 \centerline{\LARGE \bf }

\vspace{1.5truecm}

\centerline{
    {\bf G. Arias-Tamargo${}^{a,b}$} \footnote{ariasguillermo@uniovi.es}
    { \bf D. Rodriguez-Gomez${}^{a,b}$} \footnote{d.rodriguez.gomez@uniovi.es}
    { \bf J. G. Russo ${}^{c,d}$} \footnote{jorge.russo@icrea.cat}}

\vspace{1cm}
\centerline{{\it ${}^a$ Department of Physics, Universidad de Oviedo}} \centerline{{\it C/ Federico Garc\'ia Lorca  18, 33007  Oviedo, Spain}}
\medskip
\centerline{{\it ${}^b$  Instituto Universitario de Ciencias y Tecnolog\'ias Espaciales de Asturias (ICTEA)}}\centerline{{\it C/~de la Independencia 13, 33004 Oviedo, Spain.}}
\medskip
\centerline{{\it ${}^c$ Instituci\'o Catalana de Recerca i Estudis Avan\c{c}ats (ICREA)}} \centerline{{\it Pg.~Lluis Companys, 23, 08010 Barcelona, Spain}}
\medskip
\centerline{{\it ${}^d$ Departament de F\' \i sica Cu\' antica i Astrof\'\i sica and Institut de Ci\`encies del Cosmos}} \centerline{{\it Universitat de Barcelona, Mart\'i Franqu\`es, 1, 08028
Barcelona, Spain }}
\vspace{1cm}

\centerline{\bf ABSTRACT}
\medskip 

We study large charge sectors in the $O(N)$ model in $6-\epsilon $ dimensions. 
For $4<d<6$, in perturbation theory,
the quartic $O(N)$ theory has a UV stable fixed point at large $N$.
It was recently  argued that this fixed point can be described in terms of an IR fixed point of a cubic $O(N)$ model. By considering a double scaling limit of large charge and weak couplings, we compute two-point and all ``extremal" higher-point correlation functions  for large charge operators and find a precise equivalence between both pictures. Instanton instabilities are found to be exponentially suppressed at large charge. We also consider correlation function of $U(1)$-invariant meson 
operators in the $O(2N)\supset U(1)\times SU(N)$ theory, as a first step towards  tests of (higher spin) $AdS/CFT$.

\noindent

\newpage

\tableofcontents

\section{Introduction}

One of the most extensively studied quantum field theories is the vectorial $O(N)$ model with $N$ real scalar fields $\phi^i$ and
quartic interaction $g\, (\vec \phi\,^2)^2$. The theory has a number of interesting applications in $d\leq 4$ as it describes universal features of critical phenomena, including, for instance, a precise description of the second-order phase transition in the three-dimensional Ising model near criticality. 

Besides its importance as a description of systems of great physical interest, the $O(N)$ model is also of relevance in the context of the $AdS/CFT$ duality. It was conjectured  \cite{Klebanov:2002ja} that the Vasiliev higher spin theories on $AdS_{d+1}$ \cite{Fradkin:1987ks,Vasiliev:1990en,Vasiliev:1992av,Vasiliev:1995dn,Vasiliev:1999ba,Vasiliev:2003ev} are dual to the singlet sector of the $O(N)$ model in $d$ dimensions. In recent years, the duality was investigated particularly in  $d<4$. It is thus  natural to inquire how the duality could work for the $d>4$ theories.

The upper critical dimension of the quartic interaction is 4. For $d<4$ the $O(N)$ model flows to an IR fixed point which is typically strongly coupled \cite{Wilson:1971dc,Wilson:1973jj}. Despite this, several techniques have been developed to study the IR fixed point and their combination gives a qualitative picture of the physics of the model. In particular, large $N$ methods, the $\epsilon$-expansion around the upper/lower critical dimensions and, more recently, the bootstrap; have yielded a qualitative understanding of the $O(N)$ model in dimensions $2 < d < 4 $, including approximate results for certain quantities such as scaling dimensions of simple operators. 

Although  for $d>4$ the theory is 
 not renormalizable by power counting,  the large $N$ analysis as in \cite{Parisi:1975im} nevertheless suggests that the theory may still admit a UV fixed point, a scenario akin to Weinberg's asymptotic safety \cite{Weinberg:1980gg} (see also \cite{Litim:2014uca}). This opens the very interesting possibility to construct an interacting, strongly coupled, non-SUSY CFT in $d=5$. The existence  of a UV fixed point at large $N$ is consistent with the observation in \cite{Fei:2014yja}, drawing on the results in \cite{Lang:1990ni,Lang:1991kp,Lang:1992pp,Lang:1992zw,Petkou:1994ad,Petkou:1995vu}, that the continuation to negative $\epsilon$ of the standard $\epsilon$-expansion yields sensible results, at least within perturbation theory. Indeed, for certain observables such as scaling dimensions of some operators, the continuation to negative $\epsilon$ --that is, in $d=4+|\epsilon|$ dimensions-- of the expressions for generic $d$ leads to compelling results for the scaling dimensions, as they lie within the unitarity bounds \cite{Parisi:1975im,Bekaert:2011cu,Bekaert:2012ux}.

Fei, Giombi and Klebanov  \cite{Fei:2014yja} proposed a UV completion of the quartic $O(N)$ model in terms of a cubic theory with $N+1$ fields and the same $O(N)$ global symmetry. 
The cubic $O(N)$ model was studied in the past \cite{Ma:1975vn} in an $\epsilon$-expansion about its critical dimension $d=6-\epsilon$. 
 By examining the one-loop $\beta$ functions, it was found that, for $N>N_{\rm cr}$, with $N_{\rm cr}\sim 1038$, the theory has an IR fixed point. This result was reproduced in \cite{Fei:2014yja} , where it was shown that, at the IR fixed point, physical observables, such as scaling dimensions of operators, remarkably agree with their counterparts in the quartic model. 
 A three loop analysis \cite{Fei:2014xta} suggested that $N_{\rm cr}$ at $\epsilon=1$ may dramatically decrease up to $N_{\rm cr}\sim 64$,
 while four loop corrections \cite{Gracey:2015tta} seem to place the value of $N_{\rm cr}$ around $N_{\rm cr}\sim 400$.

The striking agreement between the IR fixed point of the cubic model and the UV fixed point of the quartic model is non-trivial. The quartic model is most conveniently studied upon performing a Hubbard-Stratonovich (HS) transformation \cite{Parisi:1975im,Moshe:2003xn}, which effectively converts it into a cubic model albeit with no (tree level) dynamics for the HS scalar field. One might think that the cubic model proposed in \cite{Fei:2014yja} is just the resummation of the higher-loop momentum dependence of the HS field. However, as shown in \cite{Eichhorn:2016hdi}, this is not the case, as the cubic model is at a larger universality class and it is only upon fine-tuning to a critical manifold that one can fall in the universality class of the quartic model with the same critical exponents. 

In any case, the fixed points of either theory can only be at most metastable. Indeed, it is known from long ago that the quartic theory at $d>4$ cannot admit a true fixed point \cite{Aizenman:1981du}. A sign of this is the fact that, in $d=4+\epsilon$ dimensions, the continuation of the standard perturbative fixed point lies at negative values of the coupling $g$, implying an inverted, unstable, potential. In turn, while the IR fixed point \cite{Fei:2014yja} of the cubic theory which UV-completes the quartic model 
lives at positive values of the couplings, the potential is cubic and it is therefore unbounded from below. An analysis of the exact RG in these models \cite{Percacci:2014tfa,Eichhorn:2016hdi,Kamikado:2016dvw} indeed shows that there is no fixed point. The problem was further studied 
in \cite{Giombi:2019upv}, where  instanton instabilities of the unbounded potentials in both theories were computed. In particular, it was shown how the instanton saddle points are responsible for giving  exponentially small imaginary parts to the scaling dimensions of the operators in both models.

In parallel, very recently it was appreciated that, on general grounds, the sectors of large charge under a global symmetry in a CFT enjoy special properties which make them analytically tractable (see, \textit{e.g.}   \cite{Hellerman:2015nra,Alvarez-Gaume:2016vff,Hellerman:2017efx,Alvarez-Gaume:2019biu,Hellerman:2017veg,Hellerman:2018sjf,Orlando:2019hte}, and \cite{Banerjee:2017fcx,Banerjee:2019jpw} for comparison with lattice simulations). This may be regarded as a generalization of very well known particular large charge limits extensively studied in the literature, such as the $pp$-wave limit or even large spin expansions in CFT. The key observation is that the correlation functions of operators with large charge under a global symmetry can be regarded, upon mapping to the cylinder, as a finite charge density state whose energy selects the relevant correlator. The scale of the charge density (operator charge) defines a UV cut-off, while the size of the sphere sets an IR cut-off. Then, provided one considers a large charge state which parametrically gives a large separation of the UV and IR scales, one can write an effective theory from which one can compute the desired correlation function. While this construction is universal, 
as it relies on  generic assumptions,  it has recently been realized that a microscopic derivation can also be provided. Focusing on the $O(2)$ model, one may consider large charge $n$ operators in the sector where $g$ scales as $n^{-1}$, so that $n$ acts as $\hbar$. Hence, the large charge limit acts as a ``classical limit" where $\hbar\sim n^{-1}$ is sent to zero at fixed $g\, n$. Of course, in the sense of the standard perturbative expansion in Feynman diagrams, this ``classical limit" digs into the quantum regime of the theory as it resums infinitely many Feynman diagrams. 
An efficient way to select the leading relevant Feynman diagrams is to consider a similar large $n$ limit but with fixed $g\,n^2$.
One can then explicitly compute and resum these diagrams, dubbed ``Kermit the frog" in \cite{Arias-Tamargo:2019xld}. 

The existence of this double scaling limit was first hinted for scalar theories long ago in \cite{Libanov:1994ug,Libanov:1995gh,Son:1995wz}, and very recently reconsidered in  \cite{Arias-Tamargo:2019xld,Badel:2019oxl,Watanabe:2019pdh,Arias-Tamargo:2019kfr}. Even though these studies concern scalar theories (mostly the $O(2)$ model), a similar double-scaling limit was found in $\mathcal{N}=2$ SQCD in \cite{Bourget:2018obm} and further studied in \cite{Beccaria:2018xxl,Beccaria:2018owt,Grassi:2019txd,Beccaria:2020azj} (see also  \cite{Hellerman:2017sur,Hellerman:2018xpi}). It is very interesting to note that in the large $n$ limit of $\mathcal{N}=2$ QCD the (Yang-Mills) instanton sector is exponentially suppressed \cite{Bourget:2018obm} and thus the perturbative series (or, if summed, its continuation to strong coupling) is exact.

In view of the simplifications in large charge sectors, it is natural to apply these techniques to the study of the $O(N)$ model in $d>4$. We shall see that in the sector of large charge operators 
the agreement between the description of the fixed points from the UV of the quartic theory and from the IR of the cubic theory can be explicitly checked, including the contribution of infinitely many diagrams in the standard Feynman expansion. 
In addition, just as in the $\mathcal{N}=2$ SQCD case, instanton corrections are absent. This will have the implication  that the sector of large charge operators is stable. In particular, the scaling dimensions of large charge operators do not have imaginary components (they are suppressed exponentially with the charge $n$, even at finite $N$). 
With this motivation in mind, in this paper we will  study a class of large charge operators which are in  the $[n,\,0,\cdots,\,0]$ representation of the $O(N)$ group.\footnote{As the models contain no pseudoscalar fields, in all cases, representations transform trivially under parity.}
One natural way to do this, for even $N$, is to re-write the theory so that it explicitly exhibits $U(1)\times SU(N)\subset O(2N)$. It turns out that the correlators of the $[n,\,0,\cdots,\,0]$ in $O(2N)$ are computed by correlators of $n$-th powers of the fundamental of $SU(N)$. Using this, we will compute the (purely real) scaling dimension of the operators in the $[n,\,0,\cdots,\,0]$ of $O(2N)$ from their 2-point functions and explicitly check the agreement between the cubic and quartic models. 
We will also explicitly compute a class of higher-point functions in the same double-scaling limit, in particular, obtaining results for the three-point function consistent with conformal symmetry.

In the case of even $n$, we will see that the same correlator can be computed as well by means of correlators of meson operators represented by $n$-fold symmetrized powers of the adjoint of $SU(N)$. 
This 
can be carried over to a version of the theory projected to $U(1)$-invariant states.

In order to solve the saddle-point equations, we will resort to  perturbation theory. As a consequence, strictly speaking, our results are valid in $d=6-\epsilon$ dimensions in the limit $\epsilon\rightarrow 0$. It is possible that, by including additional corrections, the results could be extrapolated to higher coupling so as to get to the region of $\epsilon=1$, that is, to the $d=5$ $O(N)$ model. This would provide a sector free of instabilities which may be of relevance in the $AdS/CFT$ context, primarily in the case of the theory projected to the $U(1)$ singlet sector.  

The rest of this paper is organized as follows. In section \ref{2-point} we introduce the class of operators that we consider and compute their 2-point functions in order to read off their scaling dimensions. We first do this in the context of the cubic model and then describe the agreement with the quartic model. We also discuss the absence of instanton contributions in the double scaling limit  which thus renders this sector stable. In section \ref{n-point} we compute higher-point functions for the so-called extremal case -- a terminology borrowed from the supersymmetric case that alludes to correlators with exactly one insertion of an antiholomorphic field. In section \ref{gaugedU(1)} we  compute the scaling dimension of meson operators.
Some concluding remarks will be made in section  \ref{conclusions},
which includes a discussion of  open problems. Finally, in the appendices we collect  some useful formulas as well as a 
standard derivation of the relevant correlation functions, including combinatorial factors, for the cubic interaction.

\section{Large charge operators in the cubic $O(2N)$ theory in $d=6-\epsilon$ dimensions}\label{2-point}

Our starting point is the $d=6-\epsilon$ dimensional theory investigated in \cite{Giombi:2019upv,Fei:2014yja}.
It is defined by the action

\begin{equation}
\label{cubictheoryUNGAUGED}
S=\int d^dx \left(\frac{1}{2} \big(\partial\vec{\varphi}\big)^2 +\frac{1}{2}\big(\partial\eta \big)^2+\frac{g_1}{2}\,\eta\,\big(\vec{\varphi}\big)^2+\frac{g_2}{6}\,\eta^3\right) .
\end{equation}
Here $\vec{\varphi}$ is a vector of $O(2N)$. As discussed in the Introduction, this theory has an IR stable fixed point for $N> N_{\rm cr}$. The critical $N_{\rm cr}$ was estimated in  \cite{Ma:1975vn,Fei:2014yja} to be $2N_{\rm cr}\sim 1038$ using the one-loop $\beta$
functions. However, further analysis  \cite{Fei:2014xta,Gracey:2015tta}
suggests that higher loop corrections may, at $\epsilon=1$, dramatically reduce this value. From the one-loop $\beta$ functions, one finds that the theory  \eqref{cubictheoryUNGAUGED} has an IR stable fixed point which at large $N$ sits at \cite{Fei:2014yja}

\begin{equation}
\label{fixedpoint}
g^*_1=\sqrt{\frac{6\,(4\pi)^3\,\epsilon}{2N}}\,\left(1+\mathcal{O}\Big(\frac{1}{N}\Big)\right)\,,\qquad g^*_2=6\,\sqrt{\frac{6\,(4\pi)^3\,\epsilon}{2N}}\,\left(1+\mathcal{O}\Big(\frac{1}{N}\Big)\right)\,.
\end{equation}
For $N<N_{\rm cr}$, the critical couplings move to the complex plane and the fixed point cannot describe a unitary theory.

\subsection{Correlation functions for a class of large charge operators} \label{cohh}

In the theory \eqref{cubictheoryUNGAUGED}, the elementary fields $\varphi^i$ fill a vector representation of $O(2N)$, whose Dynkin labels are $[1,\,0,\cdots,\, 0]_{D_N}$. Composite operators are then formed from their products and derivatives. Let us consider the class of operators formed solely by symmetrized (as we are dealing with bosons) powers of the $\varphi^i$. One can check that

\begin{equation}
\label{symmetrizedproduct}
{\rm Sym}^n\Big([1,\,0,\cdots,\,0]_{D_N}\Big)=\sum_{i=0}^{\lfloor \frac{n}{2}\rfloor}\,[n-2i,\,0,\cdots,\,0]_{D_N}\,.
\end{equation}
It is clear that all but the $i=0$ term in the sum in \eqref{symmetrizedproduct} will contain powers of $\vec{\varphi}\,^2$ when constructed in terms of fields. As an illustrative example, one may consider the case of even $n$, when the last term in the sum in \eqref{symmetrizedproduct} is $[0,\,0,\cdots,\,0]_{D_N}$; corresponding to $(\vec{\varphi}\,^2)^{\frac{n}{2}}$. For reasons which will become clear momentarily, let us consider precisely the $i=0$ term in the sum  in  \eqref{symmetrizedproduct}, \textit{i.e.} the $[n,\,0\cdots 0]_{D_N}$ representation of $O(2N)$.\footnote{Let us stress that our $[n,\,0\cdots 0]_{D_N}$ operator is composed out of $n$ fields and thus has --classical-- scaling dimension $\Delta_{\rm cl}=n\,(2-\frac{\epsilon}{2})$ in $d=6-\epsilon$. Of course, one may construct operators in the same representation by adding arbitrary powers of the singlet $|\vec{\varphi}|^2$, increasing arbitrarily its dimension.} Now, $O(2N)$ has a $SU(N)\times U(1)$ subalgebra. When branched in $SU(N)\times U(1)$, the $[n,\,0\cdots 0]_{D_N}$ gives

\begin{equation}
\label{branching}
[n,\,0,\cdots,\, 0]_{D_N}\rightarrow \sum_{k=0}^n [n-k,\,0\cdots 0,\,k]_{\mathbf{(n-2\,k)}}\,,
\end{equation}
where the RHS is meant to refer to $A_{N-1}$ Dynkin labels and the subscript is the $U(1)$ charge. 

Let us consider the $k=0$ term in the sum on the RHS of \eqref{branching}. In terms of fields, this representation is  easily constructed:  in terms of the complex combinations $\phi^I=\frac{\varphi^I+i\,\varphi^{I+N}}{\sqrt{2}}$ with $I=1,\cdots,\,N$, the action of the theory is

\begin{equation}
\label{cubictheory}
S=\int d^dx \left( |\partial\vec{\phi}|^2 +\frac{1}{2}\big(\partial\eta\big)^2+g_1\,\eta \,|\vec{\phi}|^2+\frac{g_2}{6}\,\eta ^3\right) ,
\end{equation}
so that

\begin{equation}
[n,\,0,\cdots,\,0]_{\mathbf{(n)}}=\phi^{I_1}\cdots\phi^{I_n}\,.
\end{equation}

Consider now the operator $\mathcal{O}_n=(\phi^1)^n$. It has $n$ indices and it has $U(1)$ charge $n$. Such operator can only be an entry of the $[n,\,0,\cdots,\,0]_{\mathbf{(n)}}$ representation corresponding to $k=0$ in the sum in \eqref{branching}. Moreover, since this operator does not contain any power of $|\vec{\phi}|^2=\vec{\varphi}\,^2$, it can only correspond to the $i=0$ term in the sum in eq. \eqref{symmetrizedproduct}, that is, to the $[n,\,0,\cdots,\, 0]_{D_N}$ representation of $O(2N)$. Thus, the operator $\mathcal{O}_n$ can only be an entry of the $[n,\,0,\cdots,\, 0]_{D_N}$ representation of $O(2N)$ with classical scaling dimension $\Delta_{\rm cl}=n\,(2-\frac{\epsilon}{2})$. In particular, it follows that, by computing correlators of $\mathcal{O}_n$, we determine the correlators of the $[n,\,0,\cdots,\, 0]_{D_N}$ representation of $O(2N)$. 

Thus, all in all, we will be interested on correlators of $\mathcal{O}_n$, from which we will read-off the correlators (and, in particular, the anomalous dimension) of the $[n,\,0,\cdots,\, 0]_{D_N}$ in $O(2N)$. Let us stress that there is no other operator to which $\mathcal{O}_n$ can correspond to, other than the $[n,\,0,\cdots,\, 0]_{D_N}$ of $O(2N)$, and thus, when computing correlators, there is no mixing to take into account.

The correlators of interest can be computed by the path integral

\begin{equation}
\langle \mathcal{O}_n(x_1)\,\overline{\mathcal{O}_n}(x_2)\rangle= Z^{-1} \int D\phi\, e^{-S_{\rm eff}}\,,
\label{correla}
\end{equation}
with $Z=\int  D\phi\, e^{-S}\ $  and

\begin{equation}
S_{\rm eff}=\int d^dx \left( |\partial\vec{\phi}|^2 +\frac{1}{2}\big(\partial\eta \big)^2+g_1\,\eta \,|\vec{\phi}|^2+\frac{g_2}{6}\,\eta ^3-n\,\log(\phi_1)\,\delta(x-x_1)-n\,\log(\phi_1^*)\,\delta(x-x_2)\right)\,.
\nonumber
\end{equation}
It is convenient to extract an overall factor of $n$ and re-scale fields and couplings as

\begin{equation}
\label{escali}
\phi^I=\sqrt{n}\,\Phi^I\,,\qquad \eta =\sqrt{n}\,\rho\,\qquad g_1=\frac{h_1}{\sqrt{n}}\,,\qquad g_2=\frac{h_2}{\sqrt{n}}\,;
\end{equation}
so that, up to a constant

\begin{equation}
\label{Seff}
S_{\rm eff}=n\,\int  d^dx \left(|\partial\vec{\Phi}|^2 +\frac{1}{2}\big(\partial\rho\big)^2+h_1\,\rho\,|\vec{\Phi}|^2+\frac{h_2}{6}\,\rho^3-\log(\Phi_1)\,\delta(x-x_1)-\log(\Phi_1^*)\,\delta(x-x_2)\right)  .
\end{equation}
Since $S_{\rm eff}$ is a function of $(n,\,h_1,\,h_2)$, when inserted in \eqref{correla}, the correlator of interest will be a function of these same variables, and hence must admit a double expansion in $n$, $h_i=g_i\,\sqrt{n}$ (this is similar to the double expansion discussed in \cite{Badel:2019oxl} for $|\phi|^4$ theory). In particular, the scaling dimension of the operator $[n,\,0,\cdots,\, 0]_{D_N}$ must have the form

\begin{equation}
\label{generalDelta}
    \Delta_{[n,\,0,\cdots,\, 0]_{D_N}}=n\,\sum_{k=0}\,n^{-k}\,\Delta_k(h_1,\,h_2)\,.
\end{equation}

Note that, in \eqref{Seff}, $n$ plays the role of $\hbar^{-1}$, and it is thus the loop counting parameter. Written in this form, it is natural to consider the ``classical limit"

\begin{equation}
n\rightarrow \infty\,,\qquad h_{1,\,2}\equiv {\rm fixed}\,.
\end{equation}
In this limit, the saddle point approximation becomes exact, and, in particular, selects the term $k=0$ in \eqref{generalDelta}. Let us stress that, even though this is formally a classical limit, what plays the role of $\hbar$ is $n^{-1}$. This ``classical limit" still resums an infinite series of Feynman diagrams in the conventional loop expansion (see Fig. \ref{diagrams}).

The saddle-point equations are given by

\begin{equation}
\label{I>1}
\partial^2\Phi_I=h_1\,\rho\,\Phi_I\,, \qquad \partial^2\Phi^*_I=h_1\,\rho\,\Phi^*_I\,,\qquad I=2,\,\cdots,\,N\, ,
\end{equation}

\begin{equation}
\label{Phi1}
\partial^2\Phi_1+\frac{1}{\Phi_1^*}\,\delta(x-x_2)=h_1\,\rho\,\Phi_1\,,\qquad \partial^2\Phi^*_1+\frac{1}{\Phi_1}\,\delta(x-x_1)=h_1\,\rho\,\Phi^*_1\, ,
\end{equation}

\begin{equation}
\label{rho}
\partial^2\rho=h_1\,|\vec{\Phi}|^2+\frac{h_2}{2}\,\rho^2\,.
\end{equation}
In the saddle-point method, the correlator is then determined as usual by the evaluation of the integrand of \eqref{correla} on the solution to the saddle-point equations. 

To further proceed, let us consider the regime where $h_1,\ h_2\ll 1$, so that we can solve \eqref{I>1}, \eqref{Phi1}, \eqref{rho} in perturbation theory. 
To begin with, note that $\Phi_I=\Phi_I^*=0$ for $I\ne 1$ is an exact solution. 
To order zero,  \eqref{Phi1}, and \eqref{rho} are solved by

\begin{eqnarray}
&&\Phi^{(0)}_1=\frac{G(x-x_2)}{\sqrt{G(x_1-x_2)}}\,,\qquad \Phi^{(0)*}_1=\frac{G(x-x_1)}{\sqrt{G(x_1-x_2)}}\,,\\ \nonumber \\
&&\rho^{(0)}=0\,,
\end{eqnarray}
where $G(x-y)$ is the the Green's function for the laplacian defined as

\begin{equation}
\partial^2G(x-y)=-\delta(x-y)\, .
\end{equation}
In the solution, we have set to one an arbitrary multiplicative constant in $\Phi^{(0)}_1$, and the inverse constant in $\Phi^{(0)*}_1$ , as they cancel out in the computation of the action. 

To the next order, given that only $\Phi_{1}^{(0)}$, $\Phi_{1}^{(0)*}$ are non-zero, \eqref{rho} gives 

\begin{equation}
\label{rho1}
\partial^2\rho^{(1)}-\frac{h_2}{2}\,\rho^{(1)}\,^2=h_1\,\frac{G(x-x_1)\,G(x-x_2)}{G(x_1-x_2)}\,.
\end{equation}
It obviously follows that $\rho^{(1)}$ is of order $h_{1}$ itself. Hence, the RHS of  \eqref{Phi1} will be of order $\mathcal{O}(h_1^2)$,
which in turn show that both $\Phi_{1}^{(1)}$ and $\Phi_{1}^{(1)*}$ will be of order $\mathcal{O}(h_1^2)$. Thus, to leading order in $h_{1,2}$, $\Phi_{1}^{(1)}=\Phi_{1}^{(1)*}=0$, and we only need to solve \eqref{rho1}. To that matter, let us introduce $\rho^{(1)}=h_1\,\varrho$. Then the equation becomes

\begin{equation}
\partial^2\varrho-h_1\,h_2\,\varrho^2=\frac{G(x-x_1)\,G(x-x_2)}{G(x_1-x_2)}\,.
\end{equation}
In perturbation theory, we can approximate this equation by

\begin{equation}
\partial^2\varrho=\frac{G(x-x_1)\,G(x-x_2)}{G(x_1-x_2)}\,.
\end{equation}
The solution to this equation is simply

\begin{equation}
\varrho=-\frac{1}{G(x_1-x_2)}\,\int d^6x_3\,G(x-x_3)\,G(x_3-x_1)\,G(x_3-x_2)\,.
\end{equation}

Therefore, the solution in perturbation theory to order $\mathcal{O}(h_i^2)$ is

\begin{eqnarray}
&&\Phi_1=\frac{G(x-x_2)}{\sqrt{G(x_1-x_2)}}+\mathcal{O}(h_i^2)\,,\qquad \Phi_1^*=\frac{G(x-x_1)}{\sqrt{G(x_1-x_2)}}+\mathcal{O}(h_i^2)\, ,\\ \nonumber \\
&& \Phi_I=\Phi_I^*=0\,,\qquad I=2,\,\cdots,\,N\, ,\\ \nonumber \\
&&\rho=-\frac{h_1}{G(x_1-x_2)}\,\int d^6x_3\,G(x-x_3)\,G(x_3-x_1)\,G(x_3-x_2)+\mathcal{O}(h_i^2)\,.
\end{eqnarray}

In order to compute the correlator, it only remains to evaluate the action on the saddle point solution. Let us  write $S_{\rm eff}=S_{\rm free}+S_{\rm int}$, with

\begin{equation}
    S_{\rm free}=n\,\int  d^dx \left(|\partial\vec{\Phi}|^2 +\frac{1}{2}\big(\partial\rho\big)^2-\log(\Phi_1)\,\delta(x-x_1)-\log(\Phi_1^*)\,\delta(x-x_2)\right)  ,
\end{equation}
and

\begin{equation}
    S_{\rm int}=n\,\int  d^dx \left(h_1\,\rho\,|\vec{\Phi}|^2+\frac{h_2}{6}\,\rho^3\right)  .
\end{equation}
Computing the free part, up to multiplicative constant, we find the expected factor $e^{-S_{\rm free}}=G(x_1-x_2)^n\sim |x_1-x_2|^{-2\,\Delta_{\rm cl}}$. Therefore

\begin{equation}
\label{correla2}
    \langle \mathcal{O}_n(x_1)\,\overline{\mathcal{O}}_n(x_2)\rangle\sim \frac{e^{-S_{\rm int}}}{|x_1-x_2|^{2\,\Delta_{\rm cl}}}\,.
\end{equation}
It only remains to compute the interaction piece of the action evaluated on the saddle point solution. We obtain

\begin{equation}
\label{Sint}
-S_{\rm int}=\frac{n\,h_1^2}{G(x_1-x_2)^2}\int d^6x\int d^6y\,G(x-x_1)\,G(x-x_2)\,G(x-y)\,G(y-x_1)\,G(y-x_2)\,.
\end{equation}
Upon shifting $x\rightarrow -x+x_1$ and $y\rightarrow -y+x_1$, this becomes ($z=x_1-x_2$)

\begin{equation}
\label{Sint}
-S_{\rm int}=\frac{n\,h_1^2}{G(z)^2}\int d^6x\int d^6y\,G(x)\,G(x-z)\,G(x-y)\,G(y)\,G(y-z)\,.
\end{equation}
$S_{\rm int}$ involves the integral $I=\int d^6x\int d^6y\,G(x)\,G(x-z)\,G(x-y)\,G(y)\,G(y-z)$, which can be written as

\begin{equation}
I=\int\frac{d^6P}{(2\pi)^6}\,e^{-i\,P\,z}\,\tilde{I}\,,\qquad \tilde{I}=\prod_{i=1}^2\,\int \frac{d^6 p_i}{(2\pi)^6}\,\tilde{G}(p_1)\,\tilde{G}(p_2)\,\tilde{G}(p_1-p_2)\,\tilde{G}(p_1+P)\,\tilde{G}(p_2+P)\,.
\end{equation}
where $\tilde G(p)=1/p^2$. 
Two-loop integrals of this form
have been computed in \cite{Grozin:2003ak}. In $d=6-\epsilon$ dimensions one finds

\begin{equation}
\tilde{I}=\frac{\pi^{6-\epsilon}}{(2\pi)^{12}}\,(P^2)^{1-\epsilon}\left[-\frac{1}{3\,\epsilon^2}-\frac{3-\gamma_E}{3\,\epsilon}+{\rm finite}\right]\,.
\end{equation}
Fourier-transforming and using the explicit expression for the Green's function (see appendix A), we finally find

\begin{equation}
\label{I}
I=\frac{1}{64\,\pi^3}\,G^2(z)\,\log|z|^2\,.
\end{equation}
Thus

\begin{equation}
-S_{\rm int}=\frac{1}{64\,\pi^3}\,n\,h_1^2\,\log|x_1-x_2|^2\,.
\end{equation}
Using this result and the value of $h_1$ at the fixed point coming from \eqref{fixedpoint}, \eqref{escali},  to leading order in $\frac{1}{N}$ we find

\begin{equation}
-S_{\rm int}=\frac{3\,\epsilon\,n^2}{N}\log|x_1-x_2|^2\,.
\end{equation}
Then \eqref{correla2} becomes

\begin{equation}
\label{correla3}
    \langle \mathcal{O}_n(x_1)\,\overline{\mathcal{O}}_n(x_2)\rangle\sim \frac{1}{|x_1-x_2|^{2\,(\Delta_{\rm cl} + \gamma_{[n,\,0,\cdots,\, 0]_{D_N}} ) }}\,, 
\end{equation}
where the anomalous dimension is given by

\begin{equation}
\label{anomalousdimension}
\gamma_{[n,\,0,\cdots,\, 0]_{D_N}}=-\frac{3\,\epsilon\,n^2}{N}\,.
\end{equation}

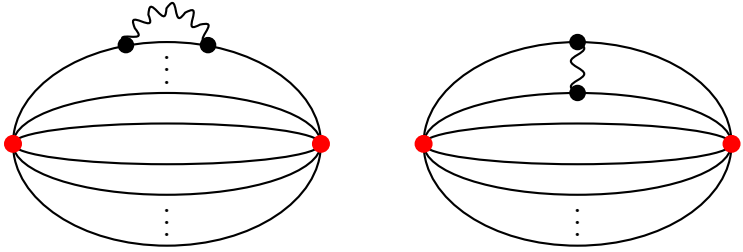
\begin{figure}[h!]
\centering
\begin{tikzpicture}[scale=1.35]
    \draw[thick] (-2,0) ellipse (1.5 and 1);
    \draw[thick] (-2,0) ellipse (1.5 and 0.5);
    \draw[thick] (-2,0) ellipse (1.5 and 0.2);
    \draw[thick,fill,red] (-3.5,0) circle (2.2pt);
    \draw[thick,fill,red] (-0.5,0) circle (2.2pt);
    \draw[thick,photon] (-2.4,0.97) .. controls (-2.2,1.35) and (-1.8,1.35).. (-1.57,0.85); 
    \draw[thick,fill] (-2.4,0.97) circle (2pt);
    \draw[thick,fill] (-1.6,0.97) circle (2pt);
    \node at (-2,0.82) {$\vdots$};
    \node at (-2,-0.68) {$\vdots$};
    
    \draw[thick] (-2+4,0) ellipse (1.5 and 1);
    \draw[thick] (-2+4,0) ellipse (1.5 and 0.5);
    \draw[thick] (-2+4,0) ellipse (1.5 and 0.2);
    \draw[thick,fill,red] (-3.5+4,0) circle (2.2pt);
    \draw[thick,fill,red] (-0.5+4,0) circle (2.2pt);
    \draw[thick, photon] (-2+4,1) -- (2,0.5); 
    \draw[thick,fill] (-2+4,1) circle (2pt);
    \draw[thick,fill] (2,0.5) circle (2pt);
     \node at (2,-0.68) {$\vdots$};
\end{tikzpicture}
\caption{Leading order diagrams for the $\langle \mathcal{O}_n(x_1)\,\overline{\mathcal{O}}_n(x_2)\rangle$ correlator. Solid lines stand for $\phi$ propagators while wavy lines stand for $\rho$ propagators.}
\label{cubicdiagramsFIGURE}
\end{figure}

It is important to stress that, even though this result is to first order in perturbation theory at weak coupling in the $h_i$ and in the ``classical limit" defined by the large $n$ limit, it resums an infinite series of Feynman diagrams in the usual perturbative expansion. To see this, note that expanding $e^{-S_{\rm int}}=1-S_{\rm int}+\cdots$, we have

\begin{equation}
    \langle \mathcal{O}_n(x_1)\,\overline{\mathcal{O}}_n(x_2)\rangle=G(x_1-x_2)^n+G(x_1-x_2)^n\,(-S_{\rm int})+\cdots
\end{equation}
Using now the expression in \eqref{Sint} for the interaction and writing it in terms of the original $g_i$ couplings, this is

\begin{eqnarray}
\label{correla4}
    &&\langle \mathcal{O}_n(x_1)\,\overline{\mathcal{O}}_n(x_2)\rangle=G(x_1-x_2)^n + \\ \nonumber && \hspace{1cm} 2\,\frac{n^2}{2}\,g_1^2\,G(x_1-x_2)^{n-2}\,\int d^6x\int d^6y\,G(x-x_1)\,G(x-x_2)\,G(x-y)\,G(y-x_1)\,G(y-x_2)+\cdots\,.
\end{eqnarray}
In the second line in \eqref{correla4} we recognize precisely the diagram on the right panel in Fig. \ref{cubicdiagramsFIGURE}.\footnote{The factor of 2 in \eqref{correla4} deserves some discussion. Strictly speaking, the free correlator is $n!\,G(x_1-x_2)^n$. The NLO correction we are computing has an extra factor of 2 with respect to this $n!$, which is that in \eqref{correla4}. See appendix \eqref{combinatorics} for further discussion.}

\begin{figure}[h!]
\centering
\begin{tikzpicture}[scale=1.35]
    \draw[thick] (-2,0) ellipse (1.5 and 1);
    \draw[thick,photon] (-2,1)--(-2,-1);
    \draw[thick,fill] (-2,1) circle (2pt);
    \draw[thick,fill] (-2,-1) circle (2pt);
    \draw[thick,fill,red] (-3.5,0) circle (2.2pt);
    \draw[thick,fill,red] (-0.5,0) circle (2.2pt);
    \node at (-4,0) {$0$};
    \node at (0,0) {$z$};
    \node at (-2,1.3) {$x$};
    \node at (-2,-1.3) {$y$};
\end{tikzpicture}
\caption{The subdiagram contributing to $-S_{\rm int}$.}
\label{cubicrelevantdiagramFIGURE}
\end{figure}
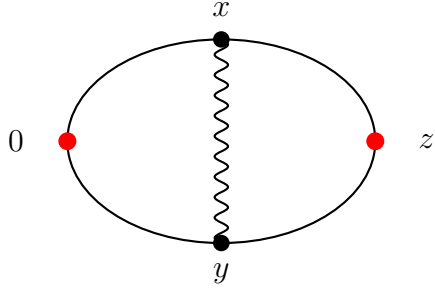

Note in particular that the overall $n^2$ stands for the combinatorics: the $\rho$ line connects each pair of $\phi$ lines, and there are $\frac{n\,(n-1)}{2}\sim \frac{n^2}{2}$ such lines, so that the diagram on the right panel in Fig. \ref{cubicdiagramsFIGURE} is proportional to $n^2\,g_1^2$. Of course, at the same order in the $g_i$'s there is the diagram on the left panel in Fig. \ref{cubicdiagramsFIGURE}, which would be proportional to $n\,g_2^2$. In our large $n$ limit with fixed $h_i$, the diagram on the left panel is suppressed and only the diagram on the right panel contributes. Moreover, it is clear that the same logic goes through for the higher order terms in \eqref{correla4}: the large $n$ limit with fixed $h_i$ limit will select the diagrams with the highest power of $n$. At weak coupling in the $h_i$'s these are the natural ``multi-ladder" generalization of the diagram on the right panel of figure \ref{cubicdiagramsFIGURE}
(see Fig. \ref{diagrams}). These infinitely many diagrams exponentiate and give the formula \eqref{correla3}, with the
 anomalous dimension given by \eqref{anomalousdimension}. To be precise, in the leading correction in the second line of \eqref{correla4}, all but the two  lines connected by $\rho$ are spectators. Thus we can factor out the $n$ free propagators, so that, for all practical purposes, the relevant diagram is that of Fig. \eqref{cubicrelevantdiagramFIGURE} --divided by $G(x_1-x_2)^2$ to factor out the overall free correlator. Then, one can easily recognize that this is precisely $-S_{\rm int}$ (\textit{c.f.} \eqref{Sint}). It is thus this subdiagram what exponentiates in the large $n$, small $h_i$ regime.

\subsection{Exact resummation in the large $n$ limit with couplings decreasing as $1/n$}\label{strict}

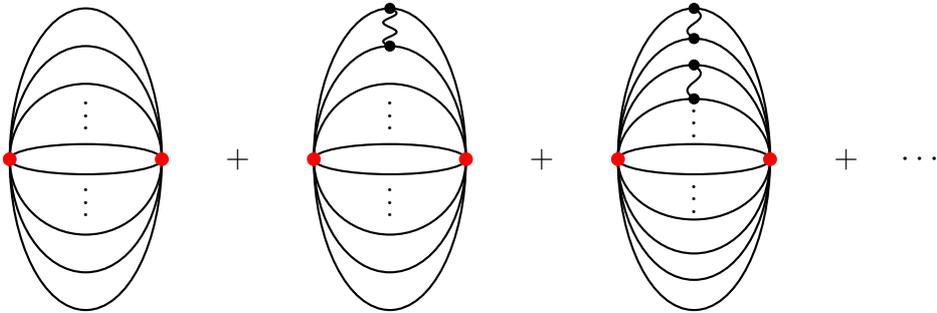
\begin{figure}[b!]
\centering
\begin{tikzpicture}

\draw[thick] (-2,0) circle (1);
\draw[thick] (-2,0) ellipse (1 and 1.5);
\draw[thick] (-2,0) ellipse (1 and 2);
\draw[thick] (-2,0) ellipse (1 and 0.2);
\node at (-2,0.7) {$\vdots$};
\node at (-2,-0.45) {$\vdots$};
\draw[thick,fill,red] (-3,0) circle (2.2pt);
\draw[thick,fill,red] (-1,0) circle (2.2pt);

\node at (0,0) {$+$};

\draw[thick] (-2+4,0) circle (1);
\draw[thick] (-2+4,0) ellipse (1 and 1.5);
\draw[thick] (-2+4,0) ellipse (1 and 2);
\draw[thick] (-2+4,0) ellipse (1 and 0.2);
\draw[photon] (2,2)--(2,1.5);
\node at (-2+4,0.7) {$\vdots$};
\node at (-2+4,-0.45) {$\vdots$};
\draw[thick,fill,red] (-3+4,0) circle (2.2pt);
\draw[thick,fill,red] (-1+4,0) circle (2.2pt);
\draw[thick,fill] (2,2) circle (1.7pt);
\draw[thick,fill] (2,1.5) circle (1.7pt);

\node at (4,0) {$+$};

\draw[thick] (6,0) ellipse (1 and 1.25);
\draw[thick] (-2+4+4,0) ellipse (1 and 0.8);
\draw[thick] (-2+4+4,0) ellipse (1 and 1.6);
\draw[thick] (-2+4+4,0) ellipse (1 and 2);
\draw[thick] (-2+4+4,0) ellipse (1 and 0.2);
\draw[photon] (2+4,2)--(2+4,1.6);
\draw[photon] (2+4,1.25)--(2+4,0.8);
\node at (-2+4+4,0.6) {$\vdots$};
\node at (-2+4+4,-0.4) {$\vdots$};
\draw[thick,fill,red] (-3+4+4,0) circle (2.2pt);
\draw[thick,fill,red] (-1+4+4,0) circle (2.2pt);
\draw[thick,fill] (2+4,2) circle (1.7pt);
\draw[thick,fill] (2+4,1.6) circle (1.7pt);
\draw[thick,fill] (2+4,1.25) circle (1.7pt);
\draw[thick,fill] (2+4,0.8) circle (1.7pt);

\node at (8,0) {$+$};
\node at (9,0) {$\cdots$};

\end{tikzpicture}
\caption{Resummation of the surviving Feynman diagrams in the large $n$ limit with fixed $\lambda_1,\ \lambda_2$.}
\label{diagrams}
\end{figure}

It is possible to consider an alternative scaling where the leading order in the perturbation series in the $h_i$'s couplings becomes exact. The correlation function \eqref{correla} can be computed again by the saddle-point method, where we define rescaled fields as $\vec\varphi = g_1^{\frac12} \vec\phi$. At the same time we define the couplings 

\begin{equation}
\lambda_1 = g_1 n\ ,\qquad \lambda_2 =g_2 n\ .    
\end{equation}
We now take the large $n$ limit, this time with $\lambda_1,\ \lambda_2$ fixed. In this limit, the saddle-point equations become

\begin{equation}
    \partial^2 \varphi_I = 0 \ ,\quad
       \partial^2 \varphi_I^* = 0 \ , \quad I=2,\cdots,N\ ,
\end{equation}
\begin{equation}
     \partial^2 \varphi_1 +\frac{\lambda_1}{\varphi_1^*}\delta(x-x_2) = 0\ ,\quad    \partial^2 \varphi_1^* +\frac{\lambda_1}{\varphi_1}\delta(x-x_1) = 0\ ,
\end{equation}
\begin{equation}
    \partial^2\eta  = | \vec \varphi|^2\ .
\end{equation}
These equations reproduce the ones obtained for the leading order in the perturbation series in section 2.1, which, with the current scaling, become exact. This is the precise analog of the limit considered in \cite{Arias-Tamargo:2019xld,Arias-Tamargo:2019kfr},
with the difference that now there is an additional field $\eta $, which mediates the interaction.  The saddle-point calculation in the double-scaling limit gives rise to the exponentiation of the second Feynman diagram of Fig. \ref{cubicdiagramsFIGURE}. The exponentiation corresponds to the full resummation of the Feynman diagrams of Fig. \ref{diagrams}, which are the only surviving Feynman diagrams in this limit, all other diagrams being suppressed by powers of $1/n$.
In the case of \cite{Arias-Tamargo:2019xld,Arias-Tamargo:2019kfr},
the relevant Feynman diagram corresponds to collapsing the $\eta $
propagator to a point, i.e. setting $x=y$ in Fig. \ref{cubicrelevantdiagramFIGURE}
(this gives rise to the ``Kermit the frog" diagram of \cite{Arias-Tamargo:2019xld,Arias-Tamargo:2019kfr}). 

In terms of the perturbation series in the $h_1,\ h_2$ couplings, organized as in section 2.1,
this limit can be understood as follows. Recall that, on general grounds, the dimension of the operator is given by \eqref{generalDelta}, where the $\Delta_k$'s are given by a perturbative series in $h_1,\ h_2$. At leading order in the $1/n$ expansion,  we can keep the $\Delta_0$ term alone.
 $\Delta_0$ admits a perturbation series expansion in the $h_1, \ h_2$'s, which  is schematically of the form (we collectively denote by $h$ both $h_{1,\,2}$)

\begin{eqnarray}
    \Delta_{[n,\,0,\cdots,\, 0]_{D_N}}
    &=&  n\,\Delta_0^{(0)}+\Delta_0^{(2)}\,n\,h^2+\Delta_0^{(4)}\,n\,h^4+\cdots \nonumber\\
    &=&  n\,\Delta_0^{(0)}+\Delta_0^{(2)}\,(\sqrt{n}\,h)^2+\frac{1}{n}\,\Delta_0^{(4)}\,(\sqrt{n}\,h)^4+\cdots \,.
\end{eqnarray}
Here $\Delta_0^{(0)}$ is the classical dimension of a scalar in $d=6-\epsilon$ dimensions, that is, $\Delta_0^{(0)}=2-\frac{\epsilon}{2}$. 

In the limit where $h_i\sqrt{n}=\lambda_i$ is kept fixed, all but the leading correction are suppressed. Making use of the
previous results, we obtain

\begin{equation}
    \Delta_{[n,\,0,\cdots,\, 0]_{D_N}}= 2\,n-\frac{\epsilon n}{2}+\Delta_0^{(2)}\,(\sqrt{n}\,h)^2=2\,n-\frac{\lambda_1^2}{64\,\pi^3} +{\cal O}(n^{-1})\,.
\end{equation}
where we have used that $\epsilon\sim n^{-2}$ on the fixed point at fixed $\lambda_{1,2}$.
Thus, the fixed $\lambda_1,\ \lambda_2$ limit reproduces the leading term of the perturbation series in  $h_1,\ h_2$, where now $h_i\sim n^{-\frac{1}{2}}$, giving $g_i\sim \frac{1}{n}$ in terms of the original couplings.

\subsection{On (the absence of) instanton contributions}

Instanton contributions are typically of order $\exp[-{\rm const.}/g_{1,2}^2]$ and are therefore exponentially suppressed at weak couplings. As the large $n$, double-scaling limit \eqref{escali} requires $g_{1,2}^2\sim 1/n$, instanton contributions will vanish exponentially as $\exp[-{\rm const.}\, n]$.
It is instructive to explicitly derive this result for the present model.

Let us first consider the equation \eqref{rho} for $\rho$. Below we will show that the term $h_1|\vec{\Phi}|^2$ on the RHS can be neglected in a self-consistent approximation at weak coupling, as it carries higher powers of the couplings. Ignoring this term, the equation becomes

\begin{equation}
\partial^2\rho=\frac{h_2}{2}\,\rho\,.
\end{equation}
It is well-known from long ago \cite{Mckane:1978me,McKane:1978md,McKane:1984eq} --and recently described in this context in \cite{Giombi:2019upv}-- that this equation admits instanton solutions $\rho_{\rm inst}$ destabilizing the theory (and giving, in particular, imaginary parts to scaling dimensions). The solution is given 

\begin{equation}
\label{rhoinst}
\rho_{\rm inst}=-\frac{12}{h_2}\,\frac{4\,\lambda^2}{(1+\lambda^2\,(\vec{x}-\vec{a})^2)^2}\,.
\end{equation}
Here $\vec{a}$ has the interpretation as the position of the instanton, while $\lambda$ corresponds to its (inverse) size, both being moduli. Since the instanton solution is of order $h_2^{-1}$, this would justify neglecting the $|\vec{\Phi}|^2$ term in the \eqref{rho} equation provided $\Phi$ is  at most
of order $h_i^0$. To check this, let us now turn to the $\Phi$-equations. Of course, $\Phi_I=\Phi_I^*=0$ for $I>1$. The remaining equations are

\begin{eqnarray}
\label{PhiInst}
&&\partial^2\Phi_1+\frac{1}{\Phi_1^*}\,\delta(x-x_2)=-\frac{h_1}{h_2}\,\frac{48\,\lambda^2}{(1+\lambda^2\,(\vec{x}-\vec{a})^2)^2}\,\Phi_1\,,\\ \nonumber \\ &&
 \partial^2\Phi^*_1+\frac{1}{\Phi_1}\,\delta(x-x_1)=-\frac{h_1}{h_2}\,\frac{48\,\lambda^2}{(1+\lambda^2\,(\vec{x}-\vec{a})^2)^2}\,\Phi^*_1\,.
\end{eqnarray}
Since $\frac{h_1}{h_2}\sim \mathcal{O}(h_i^0)$, the solution to these equations is of order $\mathcal{O}(h_i^0)$, consistently with the assumption for the instanton solution to \eqref{rho}. 

Evaluating the action on the solution,  one obtains the instanton action given by

\begin{equation}
\label{instantonactionsmalllambda}
S_{\rm inst}\sim-n\,\frac{768\,\pi^3}{5\,h_2^2}+\mathcal{O}(h_i^0)\,.
\end{equation}
Thus, in the large $n$ limit (and actually for any value of $h_i$), the instanton action goes to $-\infty$, and hence the instanton contribution is exponentially  suppressed. The absence of instanton contributions in the large $n$, double scaling limit was first noticed in the supersymmetric context in \cite{Bourget:2018obm}. Thus we conclude that, in the sector of large charge operators, instantons are completely suppressed even at finite $N$. This shows that large charge operators are free of instanton instabilities and therefore represent a stable sector with  real scaling dimensions.

\subsection{The quartic theory avatar}
\label{qavatar}

Let us consider the quartic $O(2N)$ theory with lagrangian 

\begin{equation}
\label{quartictheoryUNGAUGED}
S=\int d^dx \left(\frac{1}{2}\, \big|\partial\vec{\varphi}\big|^2 +\frac{g}{4}\,(\vec{\varphi}^2)^2 \right).
\end{equation}
This theory has a UV fixed point at negative $g$ in $4<d<6$ dimensions. It has been conjectured in \cite{Fei:2014yja} that the cubic theory \eqref{cubictheoryUNGAUGED} is a UV completion of this theory. Just as in the cubic theory, we will introduce the suitable complex combinations of fields to make explicit a $U(N)$ subgroup of the full $O(2N)$. 
The standard treatment  \cite{Parisi:1975im,Domb:1976bk} that generates the $1/N$ expansion is
by  considering a Hubbard-Stratonovich transformation, which leads to the action

\begin{equation}
\label{quarticHS}
S=\int d^dx \left(|\partial\vec{\phi}|^2 +\sigma |\vec{\phi}|^2-\frac{1}{4\,g}\,\sigma^2\right) .
\end{equation}
The original theory is recovered upon integrating out $\sigma$.
At the (now UV) fixed point the last term can be dropped, and the dynamic is just described by the action

\begin{equation}
S=\int d^dx \left(|\partial\vec{\phi}|^2 +\sigma |\vec{\phi}|^2\right) .
\end{equation}
It is convenient to re-scale $\sigma$ as in \cite{Fei:2014yja} so that  the action becomes\footnote{Recall that we are considering the $O(2N)$ model. Yet, as $\phi$ is complex, in the one-loop contribution to the $\sigma$ propagator $N$ $\phi$'s are running.}

\begin{equation}
\label{quartic}
S=\int d^dx \left( |\partial\vec{\phi}|^2 +\frac{1}{\sqrt{N}}\,\sigma |\vec{\phi}|^2\right) .
\end{equation}
A 2-point function for $\sigma$ is induced at one-loop \cite{Parisi:1975im}. In position space, it reads
(see  appendix \eqref{combinatorics}, which includes a discussion of some relevant factors; see also \cite{Fei:2014yja})
\begin{equation}
\label{sigmapropagator}
\langle \sigma(x)\,\sigma(0)\rangle=\frac{C_d}{2\,(x^2)^2}\,\qquad C_d=\frac{2^{2+d}\,\Gamma(\frac{d-1}{2})\,\sin(\frac{\pi\,d}{2})}{\pi^{\frac{3}{2}}\,\Gamma(\frac{d}{2}-2)} \,.
\end{equation}
Note that, if one substitutes $d=6-\epsilon$, one finds

\begin{equation}
\label{propagatorsigma}
\langle \sigma(x)\,\sigma(0)\rangle=\frac{1}{2}\,\tilde{C}_6\,G(x) \,;
\end{equation}
where $G(x)$ is the 6d (scalar) propagator, and

\begin{equation}
\label{C6}
\tilde{C}_6=(2\,N)\,g_1^*\,^2\,,
\end{equation}
being $g_1^*$ the value of $g_1$ at the fixed point given by \eqref{fixedpoint}. Note that $g_1^*\,^2\sim \epsilon$. Thus, to leading order in $\epsilon$, $G(x)$ in \eqref{propagatorsigma} is just the 6d propagator.

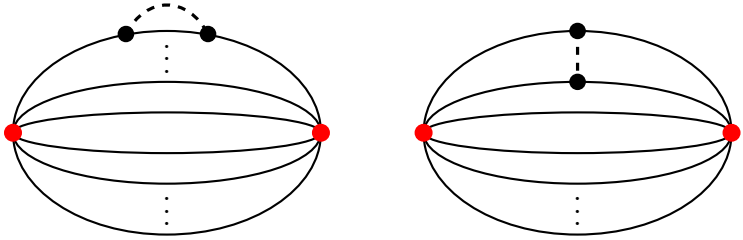
\begin{figure}[h!]
\centering
\begin{tikzpicture}[scale=1.35]
    \draw[thick] (-2,0) ellipse (1.5 and 1);
    \draw[thick] (-2,0) ellipse (1.5 and 0.5);
    \draw[thick] (-2,0) ellipse (1.5 and 0.2);
    \draw[thick,fill,red] (-3.5,0) circle (2.2pt);
    \draw[thick,fill,red] (-0.5,0) circle (2.2pt);
    \draw[very thick,dashed] (-2.4,0.97) .. controls (-2.2,1.35) and (-1.8,1.35).. (-1.6,0.97); 
    \draw[thick,fill] (-2.4,0.97) circle (2pt);
    \draw[thick,fill] (-1.6,0.97) circle (2pt);
    \node at (-2,0.82) {$\vdots$};
    \node at (-2,-0.68) {$\vdots$};
    
    \draw[thick] (-2+4,0) ellipse (1.5 and 1);
    \draw[thick] (-2+4,0) ellipse (1.5 and 0.5);
    \draw[thick] (-2+4,0) ellipse (1.5 and 0.2);
    \draw[thick,fill,red] (-3.5+4,0) circle (2.2pt);
    \draw[thick,fill,red] (-0.5+4,0) circle (2.2pt);
    \draw[very thick,dashed] (-2+4,1) -- (2,0.5); 
    \draw[thick,fill] (-2+4,1) circle (2pt);
    \draw[thick,fill] (2,0.5) circle (2pt);
     \node at (2,-0.68) {$\vdots$};
\end{tikzpicture}
\caption{Diagrams contributing to the $\langle \mathcal{O}_n(x_1)\,\overline{\mathcal{O}}(x_2)\rangle$ correlator in the quartic theory to the leading-non-trivial order. Solid lines correspond to $\phi$ fields, while dashed lines are $\sigma$ fields.}
\label{quarticdiagramsFIGURE}
\end{figure}

The quartic theory \eqref{quartictheoryUNGAUGED} exhibits the same $SU(N)$ global symmetry as the cubic theory \eqref{cubictheory}. Thus, we can consider the same $\mathcal{O}_n=(\phi^1)^n$ operator which, by the same arguments as above, can only belong to the dimension $\Delta_{\rm cl}$ operator in the $[n,\,0\cdots 0]_{D_N}$ representation of the original $O(2N)$. Thus, the correlator $\langle \mathcal{O}_n(x_1)\,\overline{\mathcal{O}}_n(x_2)\rangle$ determines its anomalous dimension. Since the $\sigma$ propagator is itself induced at one-loop, in the present formulation we do not have an easy path integral representation for the correlator. Yet, we can compute it directly in perturbation theory. The leading correction to the free theory is given by the diagrams in Fig. \ref{quarticdiagramsFIGURE}.

Just as for the cubic theory, the diagram on the left panel is suppressed with respect to the diagram on the right panel of  Fig. \ref{quarticdiagramsFIGURE} in the large $n$ limit by a factor $1/n$.  Hence, we only need to evaluate the  diagram on the right. Moreover, the combinatorics of the diagram on the right panel of Fig. \ref{quarticdiagramsFIGURE} are just as in the cubic case and thus, at large $n$, the diagram comes multiplied by $\frac{n^2}{2}$, giving (we include the aforementioned factor of $2$ which cancels the $\frac{1}{2}$ in the $\sigma$ propagator)

\begin{equation}
D=\frac{n^2}{2}\,\frac{\tilde{C}_6}{N}\,G(x_1-x_2)^{n-2}\,\int d^6x\int d^6y\,G(x-x_1)\,G(x-x_2)\,G(x-y)\,G(y-x_1)\,G(y-x_2)\,.
\end{equation}
Using the explicit value of $\tilde{C}_6$ in \eqref{C6}, we find

\begin{equation}
D=n^2\,g_1^*\,^2\,G(x_1-x_2)^{n-2}\,\int d^6x\int d^6y\,G(x-x_1)\,G(x-x_2)\,G(x-y)\,G(y-x_1)\,G(y-x_2)\,.
\end{equation}
This precisely recovers the second line in \eqref{correla4} (evaluated at the fixed point), implying a striking match with the anomalous dimension  computed from the cubic theory.

\section{Extremal higher-point functions} \label{n-point}

\subsection{The cubic theory}

Let us now consider the correlation function of an arbitrary number of operators in  representations $[n_i,\,0,\cdots,\, 0]_{D_N}$, with
$i=1,\cdots,\,k$,  
 and one operator in the conjugate representation
 $[\sum_{i=1}^kn_i,\,0,\cdots,\, 0]_{D_N}$, with all $n_i$ of order $n\gg 1$. Correlation functions of this form were dubbed {\it extremal} in \cite{Arias-Tamargo:2019kfr}. The name is taken from
 superconformal field theories, where  correlation functions of $k$ chiral primary operators and one antichiral primary operator
turn out to have a simpler structure because of supersymmetry. 
While here there is no supersymmetry, the extremal correlators are nevertheless far more simple than non-extremal correlators
\cite{Arias-Tamargo:2019kfr}.

We shall now compute the correlation function

\begin{equation}
\langle \phi_1(x_1)^{n_1}\,\cdots\,\phi_1(x_k)^{n_k}\,\overline{\phi_1}(y)^{\sum_i n_i}\rangle\,.
\end{equation}
The path integral representation of this correlator is given in terms of the action

\begin{eqnarray}
S_{\rm eff} &=& n\,\int d^dx \bigg( |\partial\vec{\Phi}|^2 +\frac{1}{2}\partial\rho^2+h_1\,\rho\,|\vec{\Phi}|^2+\frac{h_2}{6}\,\rho^3
\nonumber\\
&-&\sum_{i=1}^k a_i\,\log(\Phi_1)\,\delta(x-x_i)-\sum_{i=1}^k\,a_i\,\log(\Phi_1^*)\,\delta(x-y)\bigg)\,,
\end{eqnarray}
where we have already extracted an overall  factor of $n$ and re-scaled fields and couplings just as in section \ref{cohh}. Moreover, we have written $n_i=a_i\,n$ and we shall consider the
large $n$ limit with all $a_i$ fixed.

In the large $n$ limit with  fixed $h_{1,2}$ and $a_i$, the saddle point-approximation becomes exact. The corresponding saddle-point equations are

\begin{equation}
\label{I>1extremal}
\partial^2\Phi_I=h_1\,\rho\,\Phi_I\,, \qquad \partial^2\Phi^*_I=h_1\,\rho\,\Phi^*_I\,,\qquad I=2,\,\cdots,\,N\,,
\end{equation}

\begin{equation}
\label{Phi1extremal}
\partial^2\Phi_1+\frac{\sum_{i=1}^ka_i}{\Phi_1^*}\,\delta(x-y)=h_1\,\rho\,\Phi_1\,,\qquad \partial^2\Phi^*_1+\sum_{i=1}^k\frac{a_i}{\Phi_1}\,\delta(x-x_i)=h_1\,\rho\,\Phi^*_1\,,
\end{equation}

\begin{equation}
\label{rhoextremal}
\partial^2\rho=h_1\,|\vec{\Phi}|^2+\frac{h_2}{2}\,\rho^2\,.
\end{equation}
We follow the same procedure as in section 2.1, by 
 solving these equations in the weak $h_{1,2}$ regime. For $I>1$, one immediately has $\Phi_I=\Phi_I^*=0$. In turn, for $\Phi_1$, $\Phi_1^*$ and $\rho$ one has (again we choose some constants judiciously)

\begin{eqnarray}
&&\Phi_1=G(x-y)+\mathcal{O}(h_i^2)\,,\qquad \Phi_1^*=\sum_{i=1}^k\,a_i\,\frac{G(x-x_i)}{G(x_i-y)}+\mathcal{O}(h_i^2)\,;\\ \nonumber \\
&&\rho=-h_1\,\sum_{i=1}^k\, \frac{a_i}{G(x_i-y)}\, \int d^6z\, G(z-y)\,G(z-x_i)\,G(x-z)+\mathcal{O}(h_i^2)\,.
\end{eqnarray}
In order to compute the correlation function of interest we need to evaluate the action on this solution. Splitting $S_{\rm eff}$ in free and interaction pieces, with

\begin{equation}
S_{\rm free}=n\,\int d^dx \left( |\partial\vec{\Phi}|^2 +\frac{1}{2}\partial\rho^2-\sum_{i=1}^k a_i\,\log(\Phi_1)\,\delta(x-x_i)-\sum_{i=1}^k\,a_i\,\log(\Phi_1^*)\,\delta(x-y)\right) ,
\end{equation}
and

\begin{equation}
S_{\rm int}=n\,\int  h_1\,\rho\,|\vec{\Phi}|^2+\frac{h_2}{6}\,\rho^3\,,
\end{equation}
up to a constant, one easily gets

\begin{equation}
e^{-S_{\rm free}}=\prod_{i=1}^k\,G(x_i-y)^{n_i}\,.
\end{equation}
On the other hand, one obtains the expression

\begin{equation}
-S_{\rm int}=n\,h_1^2\,\sum_{i=1}^k\,\sum_{j=1}^k \,\frac{a_i\,a_j}{G(x_i-y)\,G(x_j-y)}\, \int d^6x\int d^6z \,G(x-y)\,G(x-x_i)\,G(z-y)\,G(z-x_j)\,G(x-z)\,.
\end{equation}
For $x_i=x_j$, the integral is the same as the integral in \eqref{Sint}, which leads to the result \eqref{I}.
More generally, we find

\begin{equation}
\int d^6x\int d^6z \,G(x-y)\,G(x-x_i)\,G(z-y)\,G(z-x_j)\,G(x-z)=\begin{cases} \frac{G(x_i-y)^2}{32\,\pi^3}\,\log |x_i-y|\,,\qquad i=j\,; \\ \\ \frac{G(x_i-y)\,G(x_j-y)}{64\,\pi^3}\,\log\frac{|x_i-y|\,|x_j-y|}{|x_i-x_j|}\,,\qquad i\ne j\,. \end{cases}
\end{equation}
Thus, at the fixed point, we have

\begin{equation}
    -S_{\rm int}=\sum_{i=1}^k\,\frac{6\,\epsilon\,n_i^2}{N}\,\log|x_i-y|+\sum_{i<j}^k\,\frac{6\,\epsilon\,n_i\,n_j}{N}\,\log\frac{|x_i-y|\,|x_j-y|}{|x_i-x_j|}\,.
\end{equation}
Therefore, we finally find

\begin{equation}
\langle \phi_1(x_1)^{n_1}\,\cdots\,\phi_1(x_k)^{n_k}\,\overline{\phi_1}(y)^{\sum_i n_i}\rangle=\frac{\mathcal{N}}{\prod_{i=1}^k|x_i-y|^{4\,n_i-\frac{6\,\epsilon\,n_i\,(\sum{n_j})}{N}}\,\prod_{i<j}|x_i-x_j|^{\frac{6\,\epsilon\,n_i\,n_j}{N}}}\, \,.
\label{genvvv}
\end{equation}
For $k=1$ this recovers the two-point correlation functions discussed above. For $k=2$, the formula \eqref{genvvv} can be neatly encoded as

\begin{equation}
\langle \phi_1(x_1)^{n_1}\,\phi_1(x_2)^{n_2}\,\overline{\phi_1}(0)^{n_1+n_2}\rangle=\frac{\mathcal{N}}{|x_1|^{\Delta_1+\bar{\Delta}-\Delta_2}\,|x_2|^{\Delta_2+\bar{\Delta}-\Delta_1}\,|x_1-x_2|^{\Delta_1+\Delta_2-\bar{\Delta}}}\, \,,
\end{equation}
which is  the expected form for the three-point function in a CFT as dictated by conformal symmetry,  with the precise expected dimensions for the operators

\begin{equation}
    \Delta_i=2\,n_i-\frac{3\,\epsilon\,n_i^2}{N}\,,\qquad \bar{\Delta}=2\,(n_1+n_2)-\frac{3\,\epsilon\,(n_1+n_2)^2}{N}\,.
\end{equation}

\subsection{The quartic theory}

\begin{figure}[t]
    \centering
\begin{tikzpicture}[scale=1.3]


\draw[thick] (-1,1) .. controls (-0.3,1.5) and (0.5,0.8) ..  (1,0);
\draw[thick] (-1,1) .. controls (-0.3,0.85) and (0.5,0.5) ..  (1,0);
\draw[thick] (-1,1) .. controls (-0.3,0.2) and (0.5,0.2) ..  (1,0);

\draw[thick] (-1,-1) .. controls (-0.3,-1.5) and (0.5,-0.8) ..  (1,0);
\draw[thick] (-1,-1) .. controls (-0.3,-0.85) and (0.5,-0.5) ..  (1,0);
\draw[thick] (-1,-1) .. controls (-0.3,-0.2) and (0.5,-0.2) ..  (1,0);

\draw[photon] (0,1)--(0,0.7);
\draw[thick,fill] (0,1.05) circle (1.8pt);
\draw[thick,fill] (0,0.65) circle (1.8pt);

\draw[thick,fill,red] (-1,1) circle (2pt);
\draw[thick,fill,red] (-1,-1) circle (2pt);
\draw[thick,fill,red] (1,0) circle (2pt);
\node at (-1.5,1) {$x_i$};
\node at (-1.5,-1) {$x_j$};
\node at (1.5,0) {$y$};
\node at (-1.5,0.15) {$\vdots$};


\draw[thick] (-1+4,1) .. controls (-0.3+4,1.5) and (0.5+4,0.8) ..  (1+4,0);
\draw[thick] (-1+4,1) .. controls (-0.3+4,0.85) and (0.5+4,0.5) ..  (1+4,0);
\draw[thick] (-1+4,1) .. controls (-0.3+4,0.2) and (0.5+4,0.2) ..  (1+4,0);

\draw[thick] (-1+4,-1) .. controls (-0.3+4,-1.5) and (0.5+4,-0.8) ..  (1+4,0);
\draw[thick] (-1+4,-1) .. controls (-0.3+4,-0.85) and (0.5+4,-0.5) ..  (1+4,0);
\draw[thick] (-1+4,-1) .. controls (-0.3+4,-0.2) and (0.5+4,-0.2) ..  (1+4,0);

\draw[photon] (0+4,0.3)--(0+4,-0.3);
\draw[thick,fill] (0+4,0.3) circle (1.8pt);
\draw[thick,fill] (0+4,-0.3) circle (1.8pt);

\draw[thick,fill,red] (-1+4,1) circle (2pt);
\draw[thick,fill,red] (-1+4,-1) circle (2pt);
\draw[thick,fill,red] (1+4,0) circle (2pt);
\node at (-1.5+4,1) {$x_i$};
\node at (-1.5+4,-1) {$x_j$};
\node at (1.5+4,0) {$y$};
\node at (-1.5+4,0.15) {$\vdots$};

\end{tikzpicture}
\caption{Types of diagrams that contribute to the extremal correlators in the cubic theory.  These coincide with the relevant diagrams in the quartic theory upon replacing the $\rho$ propagators by $\sigma$ propagators.}
\label{diagram}
\end{figure}
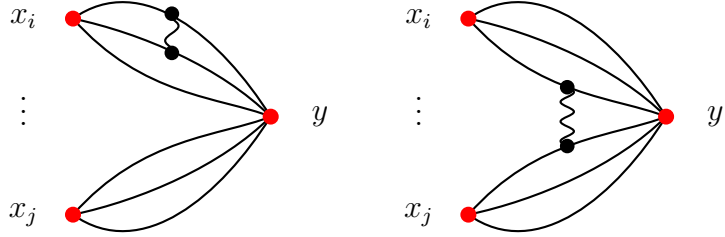

In order to compute extremal correlators in these theories, we proceed as in section \ref{qavatar} by applying the saddle-point method.
By expanding the interaction factor $e^{-S_{\rm int}}$ in powers of $S_{\rm int}$,  one can check that the diagrams contributing to the extremal correlators in the large $n$ limit are those in figure \ref{diagram}. Just as in the 2-point case, one can easily show that these are indeed the Feynman diagrams that carry  the highest power of $n$ and are thus selected in our limit.

Just as for the 2-point functions, we can compute the relevant correlation function order by order in perturbation theory. In the end, since the combinatorics is just the same as in the cubic model, the relevant diagrams are formally identical. Thus, to check agreement of the two theories it is sufficient to check the first order. The corresponding diagrams are identical to the diagrams shown in figure \ref{diagram}, upon replacing the propagator lines of the elementary field $\rho $ of the cubic model by
the induced propagator of the HS field $\sigma $ (denoted by a dashed line).
However,  \eqref{propagatorsigma}, \eqref{C6} show that, just as for the 2-point function, on the fixed point the diagrams with the exchanged scalar being $\sigma$ will be identical to the diagrams in the cubic theory with the exchanged scalar $\rho$, thus ensuring the agreement of the two computations.

\section{Correlation functions for meson operators} \label{gaugedU(1)}

The branching of the $[n,\,0,\cdots,\,0]_{D_N}$ of $O(2N)$ into $U(1)\times SU(N)$ in \eqref{branching} has an interesting particularity for even $n$. For $[2\,n,\,0,\cdots,\,0]_{D_N}$, the RHS of \eqref{branching} contains, for $k=n$, the representation $[n,\,0,\cdots,\,0,\,n]_{\mathbf{0}}$ of $U(1)\times SU(N)$. For $n=1$, from the point of view of the theory written as in \eqref{cubictheory}, such operator corresponds to a ``meson" operator $\mathcal{M}^i_j=\bar{\phi}_j\,\phi^i$ in the adjoint of $SU(N)$, which is neutral under $U(1)$. Then, higher $n$ corresponds to symmetrized $n$-fold products of this meson operator. In particular, it follows that correlators of the  $[2\,n,\,0,\cdots,\,0]_{D_N}$ can also be computed through correlators of $n$-fold symmetrized powers of meson operators.\footnote{Anomalous dimensions for similar meson operators in the quartic $O(N)$ model in $4-\epsilon$ dimensions were recently computed in \cite{Antipin:2020abu}.}

Consider now $\mathcal{M}^n=(\phi^1\,\bar{\phi}_2)^n=(\phi^1\,(\phi^2)^*)^n$. This operator clearly belongs to ${\rm Sym}^n([1,\,0\cdots 0,\,0]\otimes [0,\,0\cdots 0,\,1])$. Moreover, since that this operator does not contain any trace, it can only be an element of the $[n,\,0\cdots 0,\,n]$ $SU(N)$ representation. Thus, we may compute correlators of the $[n,\,0\cdots 0,\,n]$ representation by means of the correlator $\langle \mathcal{M}^n(x_1)\,\overline{\mathcal{M}}^n(x_2)\rangle$. Once again, $\mathcal{M}^n$ can only sit in the $[n,\,0\cdots 0,\,n]$
representation and thus there is no allowed mixing.
While we could read-off the correlators and dimensions for the  $[2\,n,\,0,\cdots,\,0]_{D_N}$ operators from the previous computation by simply replacing $n$ by $2n$, in the following we will explicitly compute the correlators of  $\mathcal{M}^n$, both as a consistency check of the previous results and also for a further application that  will be discussed in section \eqref{gauging}.

\subsection{Anomalous dimensions for mesons}

We can give a path integral representation for the two-point correlation function:

\begin{equation}
Z^{-1} \int D\phi\  \mathcal{M}^n(x_1)\,\overline{\mathcal{M}}^n(x_2)\,e^{-S}=
Z^{-1}\int D\phi\  e^{-S_{\rm eff}}\,,
\end{equation}
where now

\begin{equation}
S_{\rm eff}=\int d^dx \left(|\partial\vec{\phi}|^2 +\frac{1}{2}\partial\eta^2+g_1\,\eta\,|\vec{\phi}|^2+\frac{g_2}{6}\,\eta^3-n\,\log(\phi_1\phi_2^*)\,\delta(x-x_1)-n\,\log(\phi_1^*\phi_2)\,\delta(x-x_2)\right) .
\end{equation}
Upon performing the same re-scaling as in section 2.1, we find

\begin{equation}
S_{\rm eff}=n\,\int d^dx \left( |\partial\vec{\Phi}|^2 +\frac{1}{2}\partial\rho^2+h_1\,\rho\,|\vec{\Phi}|^2+\frac{h_2}{6}\,\rho^3-\log(\Phi_1\Phi_2^*)\,\delta(x-x_1)-\log(\Phi_1^*\Phi_2)\,\delta(x-x_2)\right) .
\end{equation}
In the double-scaling limit  $n\to \infty$  with fixed $h_i$, the saddle-point approximation becomes exact. The saddle-point equations are now given by

\begin{equation}
\partial^2\Phi_I=h_1\,\rho\,\Phi_I\,, \qquad \partial^2\Phi^*_I=h_1\,\rho\,\Phi^*_I\,,\qquad I=3,\,\cdots,\,N\, ,
\end{equation}

\begin{equation}
\partial^2\Phi_1+\frac{1}{\Phi_1^*}\,\delta(x-x_2)=h_1\,\rho\,\Phi_1\,,\qquad \partial^2\Phi^*_1+\frac{1}{\Phi_1}\,\delta(x-x_1)=h_1\,\rho\,\Phi^*_1\, ,
\end{equation}

\begin{equation}
\partial^2\Phi_2+\frac{1}{\Phi_2^*}\,\delta(x-x_1)=h_1\,\rho\,\Phi_2\,,\qquad \partial^2\Phi^*_2+\frac{1}{\Phi_2}\,\delta(x-x_2)=h_1\,\rho\,\Phi^*_2\, ,
\end{equation}

\begin{equation}
\partial^2\rho=h_1\,|\vec{\Phi}|^2+\frac{h_2}{2}\,\rho^2\,.
\end{equation}
Just as in section \ref{cohh}, we may solve these equations in perturbation theory. To order $\mathcal{O}(h_i^2)$, we find

\begin{eqnarray}
\label{U(1)sol}
&&\Phi_1=\frac{G(x-x_2)}{\sqrt{G(x_1-x_2)}}+\mathcal{O}(h_i^2)\,,\qquad \Phi_1^*=\frac{G(x-x_1)}{\sqrt{G(x_1-x_2)}}+\mathcal{O}(h_i^2)\, ,\\ \nonumber \\
&&\Phi_2=\frac{G(x-x_1)}{\sqrt{G(x_1-x_2)}}+\mathcal{O}(h_i^2)\,,\qquad \Phi_2^*=\frac{G(x-x_2)}{\sqrt{G(x_1-x_2)}}+\mathcal{O}(h_i^2)\, , \\ \nonumber \\ 
&& \Phi_I=\Phi_I^*=0\,,\qquad I=3,\,\cdots,\,N\, ,\\ \nonumber \\
&&\rho=-\frac{2\,h_1}{G(x_1-x_2)}\,\int d^6x_3\,G(x-x_3)\,G(x_3-x_1)\,G(x_3-x_2)+\mathcal{O}(h_i^2)\,.
\end{eqnarray}
The non-trivial part of the correlator is given by the interaction term of the action, which reads

\begin{equation}
-S_{\rm int}=\frac{4\,n\,h_1^2}{G(x_1-x_2)^2}\int d^6x\int d^6y\,G(x-x_1)\,G(x-x_2)\,G(x-y)\,G(y-x_1)\,G(y-x_2)\,.
\end{equation}
This involves the same integral computed before in \eqref{Sint}.
We thus obtain

\begin{equation}
-S_{\rm int}=\frac{1}{16\,\pi^3}\,n\,h_1^2\,\log|x_1-x_2|^2\,.
\end{equation}
Substituting the values of the couplings at the fixed point  \eqref{fixedpoint}, we find that the anomalous dimension of the operator is

\begin{equation}
\label{anomalousdimensionGAUGED}
\gamma_{[n,\,0\cdots 0,\,n]}=-\frac{12\,\epsilon\,n^2}{N}\,.
\end{equation}
As anticipated, this is precisely the result in \eqref{anomalousdimension} upon replacing $n$ by $2\,n$. Moreover, just as in section 2, in the large $n$ limit with fixed $h_i$  instantons are exponentially suppressed. Thus, in the $n\to\infty $ limit these operators do not have imaginary parts in their scaling dimensions.

\subsubsection*{The quartic theory}

Let us now study the correlation functions of the same $U(1)$-invariant operators
$\mathcal{M}^n,\ \overline{\mathcal{M}}^n$ 
in the quartic model \eqref{quartic}. 
The computation of the two-point correlation function turns out to be essentially identical to that of the $\mathcal{O}_n,\ \overline {\mathcal{O}}_n$ operators, with the only difference that the combinatorial factor $\mathcal{C}$
of the diagram is now slightly different. Note that our correlator is a two-point function for the meson  $\phi_1^n\,(\phi_2^*)^n$. Therefore, the tree level diagram contains $2\,n$ lines: $n$ of $\phi_1$ and $n$ of $\phi_2^*$. In this case, the diagram on the right panel of Fig. \ref{quarticdiagramsFIGURE} now consists on a $\sigma$ joining any pair of such lines. There are $\frac{2n\,(2n-1)}{2}\sim 2\,n^2$ such diagrams, so that $\mathcal{C}=2\,n^2$. Thus, borrowing the computation  from section 2, the final result is

\begin{equation}
\gamma_{[n,\,0\cdots 0,\,n]}= -\frac{12\epsilon\,n^2}{N}\,,
\end{equation}
which precisely agrees with \eqref{anomalousdimensionGAUGED}. Note that, once again, we have only computed the first correction to the anomalous dimension for large $n$ operators. However, 
using the scaling of section \ref{strict},
one can prove that  the combinatorics is such that higher order corrections exactly exponentiate \cite{Arias-Tamargo:2019xld} with  the same anomalous dimension as in \eqref{anomalousdimensionGAUGED}. 

\subsection{Projection to the $U(1)$-invariant sector through gauging}\label{gauging}

The $U(1)$-invariant sector of the vectorial $O(N)$ model is
believed to play a relevant role in $AdS/CFT$ dualities.
According to  \cite{Klebanov:2002ja}, in $d=5$ the model should have an $AdS_6$ gravity dual description in terms of the higher spin theories. 
It is therefore of interest to discuss possible ways to    project the theory to the $U(1)$-invariant sector.
As in previous sections, here we shall consider the theory in $d=6-\epsilon$ in the small
 $\epsilon>0 $ regime. Upon projecting to the $U(1)$-invariant sector, the operators $\mathcal{O}_n$ are no longer available, but the mesons remain. 
A natural way to implement dynamically the projection to the $U(1)$-invariant sector in the  cubic theory  is by  gauging the $U(1)$ symmetry. 
We start with the action \eqref{cubictheory}  
and gauge the $U(1)$ in $U(1)\times SU(N)\subset O(2N)$. Besides trading derivatives by gauge-covariant derivatives, one should add all marginal terms to the lagrangian. This gives rise to a  higher-derivative theory, which was recently studied in \cite{Khachatryan:2019veb}. Such higher-derivative theory was shown to admit three IR fixed points: in two of the fixed points the $g_{1,\,2}$ couplings attain the same value \eqref{fixedpoint} as in \cite{Fei:2014yja} (they differ in the values of other couplings which, as it will be clear below, are not important for our purposes). The third fixed point, dubbed ${\rm FP}_2$ in \cite{Khachatryan:2019veb}, corresponds to critical scalar QED and it will not be interesting for our purposes.\footnote{The one-loop $\beta$ function for general six-dimensional renormalizable models containing the Yang-Mills part was computed in \cite{Casarin:2019aqw} for a general gauge group.}

Let us consider the computation of two-point correlation functions of the meson operators in the higher-derivative gauge theory. Including the operator insertions to compute the correlators of interest and dropping the gauge-fixing term, we need to consider\footnote{The notation is slightly changed with respect to \cite{Khachatryan:2019veb}. We also tune all mass parameters to zero. These include, in particular, the standard kinetic term for the gauge field. As these parameters have  a large classical $\beta$ function, in searching for an IR fixed point, one is forced to set them to zero.} 

\begin{eqnarray}
\label{cubictheoryGAUGED}
S_{\rm eff}&=&\int d^dx  \bigg(|D\vec{\phi}|^2 +\frac{1}{2}\big(\partial\eta\big)^2+g_1\,\eta \,|\vec{\phi}|^2+\frac{g_2}{6}\,\eta ^3+\frac{1}{4}(\partial F)^2+\frac{g_3}{2}\,\eta\,F^2 \nonumber \\ \nonumber  && \\ &&   -n\,\log(\phi_1\phi_2^*)\,\delta(x-x_1)-n\,\log(\phi_1^*\phi_2)\,\delta(x-x_2)\bigg)\,,\qquad D_{\mu}=\partial_{\mu}-i\,e\,A_{\mu}\,. 
\end{eqnarray}
Upon performing the change of variables in \eqref{escali} and defining

\begin{equation}
    A_{\mu}=\sqrt{n}\,a_{\mu}\,,\qquad g_3=\frac{h_3}{\sqrt{n}}\,,\qquad e=\frac{q}{\sqrt{n}}\, ,
\end{equation}
one gets

\begin{eqnarray}
\label{cubictheoryGAUGEDRe-scaled}
S_{\rm eff}&=&n\,\int d^dx \bigg( |D\vec{\Phi}|^2 +\frac{1}{2}\big(\partial\rho\big)^2+h_1\,\rho \,|\vec{\Phi}|^2+\frac{h_2}{6}\,\rho ^3+\frac{1}{4}(\partial f)^2+\frac{h_3}{2}\,\rho\,f^2 \nonumber \\ \nonumber  && \\ &&   -\log(\Phi_1\Phi_2^*)\,\delta(x-x_1)-\log(\Phi_1^*\Phi_2)\,\delta(x-x_2)\bigg)\,; 
\end{eqnarray}
where $f=da$ and the covariant derivative is now $D_{\mu}=\partial_{\mu}-i\,q\,a_{\mu}$.

We can now take the corresponding double-scaling limit where $n\rightarrow\infty$ while $\{q,\,h_i\}$ are held fixed. Then, the computation of the correlators once again boils down to the evaluation of \eqref{cubictheoryGAUGEDRe-scaled} on the solution to the saddle point equations. These equations are now, \textit{a priori}, more complicated due to the presence of the gauge field. However, evaluating the gauge current $j^{\mu}\sim i\,(\vec{\Phi}^{\dagger}\,\partial^{\mu}\vec{\Phi}-{\rm h.c.})$ on the leading perturbative solution above, it is straightforward to see that it exactly vanishes. Therefore the configuration does not source the gauge field and thus the computation becomes identical to the 
computation of section 2. This result could have been anticipated, since the operator insertions sourcing the saddle point equations carry no electric charge (they are gauge-invariant operators), so that, at the ``classical level" (recall that the large $n$ limit is a classical limit) the gauge field is not excited. In summary, the projection to the $U(1)$-invariant sector
through gauging does not change
the two-point correlation function of mesons.

\section{Conclusions}\label{conclusions}

The $O(N)$ model with the familiar quartic potential in $4<d<6$ possesses a UV fixed point in perturbation theory. The proposed UV completion in terms of a cubic model  \cite{Fei:2014yja} leads to a perturbative IR fixed point, although it faces the expected problems of non-perturbative instabilities due to the fact that the potential is not bounded from below. In this paper we have investigated sectors of large charge operators where these instabilities are exponentially suppressed at finite $N$. In addition, 
this sector enjoys special simplifications. To begin with, these operators  do not mix with other operators. Hence, their scaling dimension can be directly read off from their 2-point functions. Moreover, correlation functions can be computed by the saddle-point method, using the same techniques as in   \cite{Arias-Tamargo:2019xld,Arias-Tamargo:2019kfr}.  In the large $n$ limit the saddle point approximation becomes exact and it resums an infinite series of Feynman diagrams.

The large $n$ limit selects the diagrams with the largest combinatorial factor. This is analogous to the Kermit-the-frog diagrams dominating the analogous limit in $d=4-\epsilon$ dimensions as described in \cite{Arias-Tamargo:2019xld}.
The main difference with respect to the quartic $O(2)$ theory discussed
in \cite{Arias-Tamargo:2019xld} is that, in the present cubic model,
the dominant Feynman diagrams contain an additional $\rho$ propagator.
However, in the relevant integration region -- which determines the logarithmic behavior as $x_1\to x_2$ -- the $\rho $ propagator is  constant, so the result of integration is essentially the same.
On the other hand, one can also understand  the matching with the UV fixed point of the quartic $O(2N)$ theory: in the latter, and upon performing a convenient Hubbard-Stratonovich transformation, the  combinatorial factor shows that the same Feynman diagrams are the dominant ones
 in the double-scaling limit. Furthermore, \eqref{propagatorsigma} together with \eqref{C6} ensure that the contribution of each diagram in the quartic theory is the same as in the cubic theory,  hence explaining the agreement between the two calculations.

The fixed point in $d>4$ (IR for the cubic theory, UV for the quartic theory) 
occurs only in  perturbation theory. This can be seen by means of the exact renormalization group, where no such extremum of the effective potential exists. The instability manifests itself through instanton corrections which give small imaginary parts to scaling dimensions. However, as we have argued, in the large charge sector  instanton instabilities are washed out: in the double scaling limit, the scaling dimensions of  large charge operators become real, as the imaginary part goes exponentially to zero. We have also discussed a double scaling limit with couplings scaling as $1/n$ --the  fixed $\lambda$ limit -- 
where our result using perturbation theory in $h_i$ becomes exact. In terms of the standard expansion in Feynman diagrams, the fixed $\lambda$ limit selects an infinite series of diagrams which can be summed with infinite radius of convergence (in the end, it recovers the exponential of the classical action). The infinite radius of convergence is  consistent with the absence of instantons in the strict $n\to\infty $ limit.
As far as the $1/n$ expansion is concerned, the existence
of instanton contributions of order $\exp(-{\rm const.}\, n)$
 indicates that the series is asymptotic.
Since instantons provide an imaginary part, we expect that the Borel transform has poles on the real axes, which occurs when the asymptotic series is not of alternate type.
It would also be  interesting to study the convergence properties of the perturbative expansion in $h_i$ of large charge correlators and its implications/relations with instanton instabilities.

In this work we have studied the large $n$ limit to first order in perturbation theory in the $h_i$'s. 
There are a number of motivations to study anomalous dimensions beyond this regime.  To begin with, it would be important to test the agreement between the quartic and the cubic theory to higher orders, where the cubic interaction in the singlet scalar field  
may give rise to new diagrams contributing to the anomalous dimensions. To leading order in $1/N$, the cubic interaction has no counterpart in the quartic model, but we expect that the effect will be
compensated by higher order corrections received by the HS propagator of the $\sigma$-field.
Moreover, the strong $h_i$ regime should be able to probe the $\epsilon\rightarrow 1$ region just as in the $d=4-\epsilon$ case discussed in \cite{Badel:2019oxl,Watanabe:2019pdh}. In this manner one may explore the large charge sector of a $d=5$ ``CFT".
Such CFT is expected  \cite{Klebanov:2002ja} to have an $AdS_6$ gravity dual description through the higher spin theories of \cite{Fradkin:1987ks,Vasiliev:1990en,Vasiliev:1992av,Vasiliev:1995dn,Vasiliev:1999ba,Vasiliev:2003ev}. In this context, the model projected to the $U(1)$ singlet is of special relevance \cite{Klebanov:2002ja}.  This projection can be implemented through  the gauging of the $U(1)$ symmetry, as done in section 4.2.
This connects to the higher-derivative theory investigated in \cite{Khachatryan:2019veb} (see also \cite{Gracey:2020kbb}). A straightforward yet very interesting extension is to consider a $U(M)$ model projected to the singlet sector with $SU(N)$ global symmetry (a model in the class of \cite{Pelissetto:2001fi}). The $U(M)$ would play the role of the $O(N)/U(N)$ symmetry in \cite{Klebanov:2002ja}, of which the singlet sector is kept, and the $SU(N)$ global symmetry would  provide a global symmetry whose large charge sector, potentially free of instabilities, may be investigated.

Methods based on large charge expansions, as the one discussed in this paper and in \cite{Arias-Tamargo:2019xld,Badel:2019oxl,Watanabe:2019pdh}, or based on the effective action (see {\it e.g.} \cite{Hellerman:2018xpi}), may be useful for a number of applications. An interesting one is
to explore non-gaussianity effects in cosmological scenarios \cite{Panagopoulos:2020sxp}. Another potential application is the study of non-conformal theories at large density (see   \cite{Orlando:2019skh,Orlando:2020yii} for related investigations), in particular  scalar QCD or brane constructions such as  \cite{Polchinski:2012nh}. Lastly,  the method used in this paper for computing correlation functions of large charge operators should also be applicable to other higher-dimensional theories (see \textit{e.g.} \cite{Gracey:2015xmw,Gracey:2016zug} for an overview). It would be very interesting to see if a double scaling limit exists in these cases
and what Feynman diagrams are resummed.

\section*{Acknowledgements}

We would like to thank S. Benvenuti  and J. Gracey for
useful comments. G.A-T and D.R-G are partially supported by the Spanish government grant MINECO-16-FPA2015-63667-P. They also acknowledge support from the Principado de Asturias through the grant FC-GRUPIN-IDI/2018/000174. G.A-T is supported by the Spanish government scholarship MCIU-19-FPU18/02221. J.G.R. acknowledges financial support from projects 2017-SGR-929, MINECO
grant FPA2016-76005-C.

\begin{appendix}

\section{Fourier transforms formul\ae}

In euclidean signature, the Fourier transform of 
$1/(x^2)^{\alpha}$ is given by

\begin{equation}
\frac{1}{(x^2)^{\alpha}}=\frac{(4\pi)^{\frac{d}{2}}\,\Gamma(\frac{d}{2}-\alpha)}{4^{\alpha}\,\Gamma(\alpha)}\,\int\frac{d^dp}{(2\pi)^d}\,\frac{e^{-ipx}}{(p^2)^{\frac{d}{2}-\alpha}}\,.
\end{equation}
As an application, it follows that the Green's function in position space is
\begin{equation}
G(x)=\frac{4^{\frac{d-2}{2}}\,\Gamma(\frac{d-2}{2})}{(4\pi)^{\frac{d}{2}}\,(x^2)^{\frac{d-2}{2}}}\,.
\end{equation}

\section{Combinatorics and propagators in the (complex) scalar  theory with cubic interaction}\label{combinatorics}

Let us consider the cubic  theory with interaction Hamiltonian

\begin{equation}
    H_{\rm int}=\int d^dx\,g\,\rho\,\bar{\phi}\,\phi\,.
\end{equation}
In the following we shall use the shorthand notation $G^{(\rho)}_{xy},\,G^{(\phi)}_{xy}$ to denote $\rho$, $\phi$ propagators from $x$ to $y$ 
($G^{(\rho,\,\phi)}_{xy}=G^{(\rho,\,\phi)}_{yx}$).

Using Dyson's formula, the expectation value of any quantity can be written as $\langle O(x)\rangle \equiv \langle O(x)\,e^{-H_{\rm int}}\rangle$ where the LHS is to be evaluated in the interacting picture (that is, for all practical purposes, with free fields).

\subsection{The 1PI diagram for the real scalar self-energy}

The (leading) 1PI diagram for $\rho$ in the cubic theory is

\begin{equation}
   \frac{1}{2}\,g^2\,\int d^dx\,\int d^dy\, \langle \rho(0)\,\rho(x)\,\bar{\phi}(x)\,\phi(x)\,\rho(y)\,\bar{\phi}(y)\,\phi(y)\,\rho(z)\rangle\,.
\end{equation}
Note that the $\frac{1}{2}$ comes from the expansion to second order of the exponential of the interacting Hamiltonian.
This gives
\begin{equation}
   \frac{1}{2}\,g^2\,\int d^dx\,\int d^dy\,G^{(\phi)}_{xy}\,G^{(\phi)}_{xy}\, \langle \rho(0)\,\rho(x)\,\rho(y)\,\rho(z)\,.
\end{equation}
Performing the $\rho$ contractions we obtain (we omit the disconnected terms)

\begin{equation}
   \frac{1}{2}\,g^2\,\int d^dx\,\int d^dy\,G^{(\phi)}_{xy}\,G^{(\phi)}_{xy}\,G^{(\rho)}_{0x}\,G^{(\rho)}_{yz}+ \frac{1}{2}\,\int d^dx\,\int d^dy\,G^{(\phi)}_{xy}\,G^{(\phi)}_{xy}\,G^{(\rho)}_{0y}\,G^{(\rho)}_{xz}\,.
\end{equation}
The two integrals are just the same and we finally find

\begin{equation}
\label{1PI}
  g^2\, \int d^dx\,\int d^dy\,G^{(\phi)}_{xy}\,G^{(\phi)}_{xy}\,G^{(\rho)}_{0x}\,G^{(\rho)}_{yz}\,.
\end{equation}

\subsection{The $\sigma$ propagator in the quartic theory}

The above discussion is of direct application to the quartic theory, where loop effects induce a 2-point function for $\sigma$ (akin to the $\rho$ in the previous discussion). The tree-level $\sigma$ propagator from the action in \eqref{quarticHS} is just $-2\,g$. Denoting the 1PI diagram by $-\Gamma$, it then follows that the $\sigma$ propagator is \cite{Parisi:1975im}

\begin{equation}
    G^{(\sigma)}=-\frac{2\,g}{1-2\,g\,\Gamma}\,.
\end{equation}
At the UV fixed point, when $g\rightarrow \infty$, $G^{(\sigma)}$ it is just the inverse of the 1PI diagram for $\sigma$.   
In momentum space,  \eqref{1PI} is given by\footnote{The signs can be checked by going back to 
Lorentzian signature, where propagators  have an extra $i$. In turn, the cubic vertex is $-\frac{i}{\sqrt{N}}$. Thus the 1PI diagram is given by $\Gamma_L$, where $\Gamma_L$ is the (Lorentzian) loop integral alone ($(-\frac{i}{N})^2$ from the vertices, $i^2$ from the $\phi$ propagators, and an overall $N$ from the $N$ $\phi$'s running in the loop). Hence $G^{(\sigma)}=-\frac{2\,i\,g}{1+2\,i\,g\,\Gamma_L}\sim \frac{1}{\Gamma_L}$. Wick-rotating to the Euclidean $\Gamma_L=i\,\Gamma$, and so $G^{(\sigma)}=i\,\frac{1}{\Gamma}$. Stripping off the $i$ to go back to Euclidean signature gives $G^{(\sigma)}=\frac{1}{\Gamma}$, with $\Gamma$ given by \eqref{xxuu}.}

\begin{equation}
\label{xxuu}
\Gamma(p)=\int \frac{d^dq}{(2\pi)^d}\,\frac{1}{(p-q)^2\,q^2}\,.
\end{equation}
Introducing Feynman parameters

\begin{equation}
\int \frac{d^dq}{(2\pi)^d}\,\frac{1}{(p-q)^2\,q^2}=\int\frac{d^dq}{(2\pi)^d}\,\int_0^1dx\,\frac{1}{(q^2+\Delta)^2}\,.\qquad \Delta=x(1-x)\,p^2\,.
\end{equation}
Computing the integral, we find

\begin{equation}
\int \frac{d^dq}{(2\pi)^d}\,\frac{1}{(p-q)^2\,q^2}=-\frac{(4\pi)^{\frac{3-d}{2}}}{2^d\,\sin\Big(\frac{d\,\pi}{2}\Big)\,\Gamma\Big(\frac{d-1}{2}\Big)}\,(p^2)^{\frac{d}{2}-2}\,.
\end{equation}
Hence

\begin{equation}
\Gamma(p)=-2\,\tilde{C}_d^{-1}\,(p^2)^{2-\frac{d}{2}}\,,\qquad  \tilde{C}_d=2^{d+1}\,(4\pi)^{\frac{d-3}{2}}\,\sin\Big(\frac{d\,\pi}{2}\Big)\,\Gamma\Big(\frac{d-1}{2}\Big)\,.
\end{equation}
Thus, in momentum space, the $\sigma$ propagator is

\begin{equation}
\langle \sigma(p)\,\sigma(-p)\rangle=\frac{\tilde{C}_d}{2\,(p^2)^{\frac{d}{2}-2}}\,.
\end{equation}
In position space this is

\begin{equation}
\langle \sigma(x)\,\sigma(0)\rangle=\frac{C_d}{2\,(x^2)^2}\,, \qquad C_d=\frac{16}{(4\pi)^{\frac{d}{2}}\,\Gamma\Big(\frac{d}{2}-2\Big)}\,\tilde{C}_d \,.
\end{equation}

\subsection{The $\langle \phi(0)^2\,\bar{\phi}^2(z)\rangle$ correlator to NLO}

Let us consider the calculation of the correlation function $\langle \phi(0)^2\,\bar{\phi}^2(z)\rangle$ to NLO order. To LO this is just the free correlator. Evaluating it through Wick contractions one easily finds

\begin{equation}
    \label{free}
    \langle \phi(0)\,\phi(0)\,\bar{\phi}(z)\,\bar{\phi}(z)\rangle = 2\,G^{(\phi)}_{0z}\,.
\end{equation}

The NLO correction comes from evaluating

\begin{equation}
    \label{NLO}
    \frac{1}{2}\,g^2\,\int d^dx\,d^dy\,\langle \phi(0)\,\phi(0)\,\rho(x)\,\bar{\phi}(x)\,\phi(x)\,\rho(y)\,\bar{\phi}(y)\,\phi(y)\,\bar{\phi}(z)\,\bar{\phi}(z)\rangle\,.
\end{equation}
This gives

\begin{equation}
    \frac{1}{2}\,g^2\,\int d^dx\,d^dy\,G^{(\rho)}_{xy}\,\langle \phi(0)\,\phi(0)\,\bar{\phi}(x)\,\phi(x)\,\bar{\phi}(y)\,\phi(y)\,\bar{\phi}(z)\,\bar{\phi}(z)\rangle\,.
\end{equation}
In order to have connected diagrams, the $\phi(0)$'s must be contracted with either the $\bar{\phi}(x)$ or the $\bar{\phi}(y)$, for which each case has two possible contractions (one for each of the two $\phi(0)$'s) thus giving

\begin{eqnarray}
   && \frac{1}{2}\,g^2\,\int d^dx\,d^dy\,G^{(\rho)}_{xy}\,2\,G^{(\phi)}_{0x}\,\langle \phi(0)\,\phi(x)\,\bar{\phi}(y)\,\phi(y)\,\bar{\phi}(z)\,\bar{\phi}(z)\rangle\\ \nonumber && +\frac{1}{2}\,g^2\,\int d^dx\,d^dy\,G^{(\rho)}_{xy}\,2\,G^{(\phi)}_{0y}\,\langle \phi(0)\,\bar{\phi}(x)\,\phi(x)\,\phi(y)\,\bar{\phi}(z)\,\bar{\phi}(z)\rangle\,.
\end{eqnarray}
By the same token, the remaining $\phi(0)$ can only be contracted either with the $\bar{\phi}(x)$ or with the remaining $\bar{\phi}(y)$, giving

\begin{equation}
    2\,g^2\,\int d^dx\,d^dy\,G^{(\rho)}_{xy}\,G^{(\phi)}_{0x}\,G^{(\phi)}_{0y}\,\langle \phi(x)\,\phi(y)\,\bar{\phi}(z)\,\bar{\phi}(z)\rangle\,.
\end{equation}
Making the final contractions, we find

\begin{equation}
   4\,g^2\,\int d^dx\,d^dy\,G^{(\rho)}_{xy}\,G^{(\phi)}_{0x}\,G^{(\phi)}_{0y}\,G^{(\phi)}_{xz}\,G^{(\phi)}_{yz}\,.
\end{equation}
Thus, all in all, to NLO, the correlator is given by

\begin{equation}
    \langle \phi(0)\,\phi(0)\,\bar{\phi}(z)\,\bar{\phi}(z)\rangle = 2\,G^{(\phi)}_{0z}+4\,g^2\,\int d^dx\,d^dy\,G^{(\rho)}_{xy}\,G^{(\phi)}_{0x}\,G^{(\phi)}_{0y}\,G^{(\phi)}_{xz}\,G^{(\phi)}_{yz}\,.
\end{equation}
We may now write this as

\begin{equation}
    \langle \phi(0)\,\phi(0)\,\bar{\phi}(z)\,\bar{\phi}(z)\rangle = 2\,\Big(G^{(\phi)}_{0z}+2\,g^2\,\int d^dx\,d^dy\,G^{(\rho)}_{xy}\,G^{(\phi)}_{0x}\,G^{(\phi)}_{0y}\,G^{(\phi)}_{xz}\,G^{(\phi)}_{yz}\Big)\,,
\end{equation}
where we have extracted the factor of 2 of the free correlator, leaving behind the relative factor of $2$ pointed out in the main text. Note that the factor of 2 stands for the 4 possible relative ways to paste the two halves of \eqref{cubicrelevantdiagramFIGURE} multiplied by the factor $\frac{1}{2}$ that arises from expanding the Dyson series. This extends in a straightforward way to the more general  case of correlators of  $\mathcal{O}_n, \ \overline{\mathcal{O}}_n$ operators, with the only difference that the overall $2$ becomes $n!$.

\end{appendix}

\end{document}